\documentclass[a4paper,onecolumn,11pt,accepted=2023-08-21]{quantumarticle}
\pdfoutput=1
\usepackage[utf8]{inputenc}
\usepackage[english]{babel}
\usepackage[T1]{fontenc}
\usepackage{amsmath, amsfonts, amssymb, amsthm}
\usepackage{hyperref}

\usepackage{cleveref}
\usepackage{thmtools}
\usepackage{thm-restate}

\usepackage{tikz}
\usepackage{pifont}
\usepackage{bbding}
\usepackage{qcircuit}

\usepackage{tikz}
\usepackage{lipsum}

\usepackage[numbers,sort&compress]{natbib}

\usepackage{graphicx, subfigure}
\usepackage{enumitem}
\usepackage{xparse}
\usepackage{physics}
\usepackage{color}

\renewcommand{\S}{\mathbf{S}}

\newcommand{\CC}{\mathbb{C}}

\newcommand{\II}{\mathbb{I}}

\newcommand{\Aa}{\mathcal{A}}
\newcommand{\Bb}{\mathcal{B}}

\newcommand{\Dd}{\mathcal{D}}

\newcommand{\Gg}{\mathcal{G}}
\newcommand{\Hh}{\mathcal{H}}
\newcommand{\Ll}{\mathcal{L}}

\newcommand{\Oo}{\mathcal{O}}

\newcommand{\Rr}{\mathcal{R}}

\newcommand{\Exp}{\mathop{\mathbf{E}\hspace{0.13em}}}

\newcommand{\inv}[1]{#1^{-1}}

\renewcommand{\Pr}{\mathop{\mathbf{Pr}\hspace{0.05em}}}

\newcommand{\good}{\mathrm{good}}

\newcommand{\eps}{\epsilon}

\newcommand{\st}{\text{ s.t. }}
\newcommand{\half}{\frac{1}{2}}
\newcommand{\pf}{\mathsf{wt}}

\newcommand{\defeq}{\mathrel{\overset{\makebox[0pt]{\mbox{\normalfont\tiny\sffamily def}}}{=}}}

\usepackage{complexity}

\renewcommand{\exp}{\mathsf{exp}}

\newcommand{\bits}{\{0,1\}}

\newcommand{\fn}[3]{#1: #2 \rightarrow #3}

\newcommand{\qubit}{\CC^{2}}
\newcommand{\qubits}[1]{(\qubit)^{\otimes #1}}

\renewcommand{\eqref}[1]{\textrm{eq.~}(\ref{#1})}

\newcommand{\YES}{\mathsf{YES}}
\newcommand{\NO}{\mathsf{NO}}

\renewcommand{\exp}{\mathrm{exp}}

\newcommand{\circled}{\tikz[baseline=(char.base)]{
    \node[shape=circle,draw,inner sep=0.15pt] (char) {\scriptsize $F$};
}}

\newcommand{\sunflower}{\textrm{\ding{96}}}
\newcommand{\isunflower}{\circled}

\newcommand{\unif}{\mathrm{unif}}
\newcommand{\ftheng}{\text{$F$ then $G$}}
\newtheorem{theorem}{Theorem}
\newtheorem{definition}[theorem]{Definition}
\newtheorem{remark}[theorem]{Remark}
\newtheorem{lemma}[theorem]{Lemma}

\newtheorem{corollary}[theorem]{Corollary}

\newtheorem{claim}[theorem]{Claim}

\newenvironment{xalign}{\subequations\align}{\endalign\endsubequations}

\begin{document}

\title{A distribution testing oracle separation between QMA and QCMA}

\author{Anand Natarajan}
\affiliation{Massachusetts Institute of Technology}
\email{anandn@mit.edu}
\thanks{This work was partially completed while a participant in the Simons Institute for the Theory of Computing program \emph{The Quantum Wave in Computing: Extended Reunion}. Natarajan thanks Elizabeth Crosson, Aram Harrow, Zhiyang He, Robin Kothari, Yupan Liu, and Mehdi Soleimanifar for helpful discussions.}

\author{Chinmay Nirkhe}
\email{nirkhe@ibm.com}
\affiliation{IBM Quantum Cambridge}
\thanks{Some of the initial ideas of this work were done while affiliated with the University of California, Berkeley. This work was partially completed while a participant in the Simons Institute for the Theory of Computing program \emph{The Quantum Wave in Computing: Extended Reunion}. Nirkhe thanks Srinivasan Arunachalam, Andrew Childs, Yi-Kai Liu, William Kretschmer, Kunal Marwaha, Umesh Vazirani, and Elizabeth Yang for helpful discussions.}

\maketitle

\begin{abstract}
 It is a long-standing open question in quantum complexity theory whether the definition of \emph{non-deterministic} quantum computation requires quantum witnesses $(\QMA)$ or if classical witnesses suffice $(\QCMA)$. We make progress on this question by constructing a randomized classical oracle separating the respective computational complexity classes. Previous separations \cite{ak-oracle,fefferman-kimmel} required a quantum unitary oracle. The separating problem is deciding whether a distribution supported on regular un-directed graphs either consists of multiple connected components (yes instances) or consists of one expanding connected component (no instances) where the graph is given in an adjacency-list format by the oracle. Therefore, the oracle is a distribution over $n$-bit boolean functions.
\end{abstract}

\section{Introduction}

There are two natural \emph{quantum} analogs of the computational complexity class $\NP$. The first is the class $\QMA$ in which a quantum polynomial-time decision algorithm is given access to a $\poly(n)$ \emph{qubit} quantum state as a witness for the statement. This class is captured by the $\QMA$-complete local Hamiltonian problem \cite{KITAEV20032} in which the quantum witness can be interpreted as the ground-state of the local Hamiltonian. The second is the class $\QCMA$ in which the quantum polynomial-time decision algorithm is given access instead to a $\poly(n)$ \emph{bit classical} state. 
While it is easy to prove that $\QCMA \subseteq \QMA$ as the quantum witness state can be immediately measured to yield a classical witness string, the question of whether $\QCMA \overset{\text{\tiny ?}}{=} \QMA$, first posed by Aharonov and Naveh \cite{quant-ph/0210077}, remains unanswered. If $\QCMA = \QMA$, then every local Hamiltonian would have an efficient classical witness of its ground energy; morally, this can be thought of as an efficient classical description of its ground state. The relevance of local Hamiltonians to condensed matter physics makes this question a central open question in quantum complexity theory \cite{open-problem-query-complexity}.

Because $\P \subseteq \QCMA \subseteq \QMA \subseteq \PSPACE$, any unconditional separation of the two complexity classes would imply $\P \neq \PSPACE$ and seems unlikely without remarkably ingenious new tools. A more reasonable goal is an oracle separation between the two complexity classes. The first oracle separation, by Aaronson and Kuperberg \cite{ak-oracle}, showed that there exists a black-box unitary problem for which quantum witnesses suffice and yet no polynomial sized classical witness and algorithm can solve the problem with even negligible success probability. A second black-box  separation was discovered a decade later by Fefferman and Kimmel \cite{fefferman-kimmel}. The Fefferman and Kimmel oracle is a completely positive trace perseving (CPTP) map called an "in-place'' permutation oracle. Both oracles \cite{ak-oracle,fefferman-kimmel} are inherently quantum\footnote{It might be reasonable to wonder if the unitary oracles can be converted into classical oracles by providing oracle access to the exponentially long classical descriptions of the respective matrices. This is not known to be true because it is unclear how to use access to the classical description to solve the $\QMA$ problem.}. Whereas, the "gold-standard'' of oracle separations --- namely black-box function separations (also known as classical oracle separations) --- only require access to a \emph{classical function} that can be queried in superposition\footnote{One reason this model is natural is that if we were given a circuit of size $C$ to implement this classical function, then we would automatically get a quantum circuit of size $C$ to implement the oracle, simply by running the classical circuit coherently. This is not true for the "in-place" permutation oracle model, assuming that one-way functions exist.}.

\subsection{Graph oracles}

The major result of this work is to prove that there exists a distribution over black-box function problems separating $\QMA$ and $\QCMA$. Each black-box function corresponds to the adjacency list of a $N \defeq 2^n$ vertex constant-degree colored graphs\footnote{A similar problem was previously conjectured to be an oracle separation for these complexity classes by Lutomirski \cite{component-mixer}.} $G = (V,E)$. Roughly speaking, a graph is a $\YES$ instance if the second eigenvalue of its normalized adjacency matrix is 1 (equivalently, if it has at least two connected components) and a graph is a $\NO$ instance if it second eigenvalue is at most $1 - \alpha$ for some fixed constant $\alpha$ (equivalently, the graph has one connected component and is expanding). We call this problem the \emph{expander distinguishing problem}. 

\paragraph{Distribution oracles} A distribution over functions (equivalently, a distribution over graphs) is a $\YES$ instance if it is entirely supported on $\YES$ graphs and a distribution over functions is a $\NO$ instance if it is entirely supported on $\NO$ graphs. 

In this work, we construct, for every $n$, families of $\YES$ and $\NO$ distributions over graphs such that following hold for the promise problem of distinguishing a graph sampled from a $\YES$ distribution from a graph sampled from a $\NO$ distribution.
\begin{enumerate}
    \item There is a $\QMA$ proof system that solves this problem, where the verifier runs in quantum polynomial time and has black-box query access to the sampled graph, and the honest prover's (quantum) witness depends only on the distribution, not on the specific sample.
    \item No $\QCMA$ proof system can solve this problem, provided the prover's (classical) witness is only allowed to depend on the distribution, and not on the sample.
\end{enumerate}

Our work is not the first to consider oracles that sample from distributions over functions. The in-place oracle separation  of~\cite{fefferman-kimmel} between $\QMA$ and $\QCMA$ used oracles that sampled random permutations. For a somewhat different problem, of separating bounded-depth quantum-classical circuits, \cite{ags22} introduced a related notion called a "stochastic oracle"---the main difference between this and our model is that a stochastic oracle resamples an instance every time it is queried.

\begin{figure}[t]
    \centering
    \begin{tabular}{c | c | c}
   \textbf{Authors}  & \textbf{Separating black box object} & \textbf{Proof techniques used}  \\ \hline \hline
   Aaronson \& Kuperberg \cite{ak-oracle}  & $n$-qubit unitaries & Adversary method \\ \hline
   Fefferman \& Kimmel \cite{fefferman-kimmel} & $n$-qubit CPTP maps & $\substack{\text{Combinatorial argument,} \\ \text{Adversary method}}$ \\
   \hline
   This work & $\substack{\text{Distributions over }n\text{-bit} \\ \text{boolean functions}}$  & $\substack{\text{Combinatorial argument,} \\ \text{Adversary method,} \\ \text{Polynomial method}}$ \\ \hline
   Conjectured & $n$-bit boolean function & ?
\end{tabular}
    \caption{List of known oracle separations}
    \label{fig:list-separations}
\end{figure}

\paragraph{Comparison with previous oracle separations between $\QMA$ and $\QCMA$}
Figure \ref{fig:list-separations} summarizes our work in relation to previous oracle separations. In terms of results, we take a further step towards the standard oracle model---all that remains is to remove the randomness from our oracle. In terms of techniques, we combine the use of counting arguments and the adversary method from previous works with a $\BQP$ lower bound for a similar graph problem, due to~\cite{ambainis-childs-liu}. This lower bound was shown using the polynomial method. We view the judicious combination of these lower bound techniques---as simple as it may seem---as one of the conceptual contributions of this paper.

\paragraph{Intuition for hardness} The expander distinguishing problem is a natural candidate for a separation between $\QMA$ and $\QCMA$ because it is an "oracular" version of the sparse Hamiltonian problem, which is complete for $\QMA$~\cite[Problem H-4]{listofqmacompleteproblems}. To see this, we recall some facts from spectral graph theory. The top eigenvalue of the normalized adjacency matrix $A$ for regular graphs is always 1 and the uniform superposition over vertices is always an associated eigenvector. If the graph is an expander (the $\NO$ case of our problem), the second eigenvalue is bounded away from $1$, but if the graph is disconnected (the $\YES$ case of our problem), then the second eigenvalue is exactly $1$. Thus, our oracle problem is exactly the problem of estimating the minimum eigenvalue of $\II-A$ (a sparse matrix for a constant-degree graph), on the subspace orthogonal to the uniform superposition state. Viewing $\II-A$ as a sparse Hamiltonian, we obtain the connection between our problem and the sparse Hamiltonian problem.

One reason to show oracle separations between two classes is to provide a \emph{barrier} against attempts to collapse the classes in the "real" world. We interpret our results as confirming the intuition that any $\QCMA$ protocol for the sparse Hamiltonian must use more than just black-box access to entries of the Hamiltonian: it must use some nontrivial properties of the ground states of these Hamiltonians. In this sense, it emulates the original quantum adversary lower bound of \cite{BBBV} which showed that any $\BQP$-algorithm for solving $\NP$-complete problems must rely on some inherent structure of the $\NP$-complete problem as $\BQP$-algorithms cannot solve unconstrained search efficiently.

\paragraph{Naturalness of the randomized oracle model}

Some care must be taken whenever one proves a separation in a ``nonstandard" oracle model---see for instance the ``trivial" example in~\cite{Aaronson08} of a randomized oracle separating $\mathsf{MA}_1$ from $\mathsf{MA}$. We believe that our randomized oracle model is natural for several reasons. Firstly, as mentioned above, randomization was used in the quantum oracle of \cite{fefferman-kimmel} for essentially the same reason: to impose a restriction on the witnesses received from the prover. Secondly, it is consistent with our knowledge that our oracle separates $\QMA$ from $\QCMA$ even when the randomness is removed (and indeed we conjecture this is the case, as described below.) Thirdly, the randomization still gives the prover access to substantial information about the graph: in particular, the prover knows the full connected component structure of the graph. As we show, this information is enough for the prover to give a \emph{quantum} witness state, that in the YES case convinces the verifier with certainty. Our result shows that even given full knowledge of the component structure, the prover cannot construct a convincing classical witness---we believe this sheds light on how a $\QMA$ witness can be more powerful than a $\QCMA$ witness.

\subsection{Overview of proof techniques}

\paragraph{Quantum witnesses and containment in oracular $\QMA$} A quantum witness for any $\YES$ instance graph is any eigenvector $\ket{\xi}$ of eigenvalue 1 that is orthogonal to the uniform superposition over vertices. The verification procedure is simple: project the witness into the subspace orthogonal to the uniform superposition over vertices, and then perform one step of a random walk along the graph, by querying the oracle for the adjacency matrix in superposition. Verify that the state after the walk step equals $\ket{\xi}$. This is equivalent to a 1-bit phase estimation of the eigenvalue. If a graph is a $\NO$ instance, then there does not exist any vector orthogonal to the uniform superposition (the unique eigenvector of value 1) that would pass the previous test. 

Whenever, the graph has a connected component of $S \subsetneq V$, then an eigenvector orthogonal to the uniform superposition of eigenvalue 1 exists.
When $\abs{S} \ll N$, this eigenvector is very close to $\ket{S}$, the uniform superposition over basis vectors $x \in S$. Notice that this state only depends on the connected component $S$ and not the specific edges of the graph.
Furthermore, the state $\ket{S'}$ for any subset $S'$ that approximates $S$ forms a witness that is accepted with high probability. 

\paragraph{Lower bound on classical witnesses} The difficulty in this problem lies in proving a \emph{lower bound} on the ability for classical witnesses to distinguish $\YES$ and $\NO$ instances. 
To prove a lower bound, we argue that any quantum algorithm with access to a polynomial length classical witness must make an exponential number of (quantum) queries to the adjacency list of the graph in order to distinguish $\YES$ and $\NO$ instances. This, in turn, lower bounds the time complexity of any $\QCMA$ algorithm distinguishing $\YES$ and $\NO$ instances but is actually slightly stronger since we don't consider the computational complexity of the algorithm between queries.

Proving lower bounds when classical witnesses are involved is difficult because the witness could be based on any property of the graph. For example, the classical witness could describe cycles, triangles, etc. contained in the graph --- while it isn't obvious why such a witness would be helpful, proving that any such witness is insufficient is a significant challenge.
One way to circumvent this difficulty is to first show a lower bound \emph{assuming} some structure about the witness\footnote{Assuming structure about a witness is a common technique in theoretical computer science and in particular lower bounds for classical witnesses of quantum statements. For example, lower bounds against natural proofs \cite{lower-bounds-natural-proofs}. Another example is the NLTS statement \cite{anshu2022nlts} which is about lower bounds for classical witnesses for the ground energy of a quantum Hamiltonian of a particular form: constant-depth quantum circuits.}, and then "remove the training wheels" by showing that the assumption holds for any good classical witness. 

\paragraph{Lower bound against "subset witnesses"} 
One structure we can assume is that the witness only depends on the set of vertices contained in the connected component $S$. This is certainly the case for the quantum witness state in \eqref{eq:intro-ideal-classical-pf}. Our result shows that any polynomial-length witness only depending on the vertices in $S$ requires an exponential query complexity to distinguish $\YES$ and $\NO$ instance graphs. 

The starting point for this statement is the exponential query lower bound \emph{in the absence of a witness} (i.e. for $\BQP$) for the expander distinguishing problem proven by Ambainis, Childs and Liu \cite{ambainis-childs-liu}, using the polynomial method. In \cite{ambainis-childs-liu}, the authors define two distributions over constant-degree regular colored graphs: the first is a distribution $P_1$ over random graphs with overwhelming probability of having a second normalized eigenvalue at most $1 - \eps_0$. The second is a distribution $P_\ell$ over random graphs with overwhelming probability of having $\ell$ connected components. Since, almost all graphs in $P_1$ are $\NO$ graphs and all graphs in $P_\ell$ are $\YES$ graphs, any algorithm distinguishing $\YES$ and $\NO$ instances must be able to distinguish the two distributions.
We first show that a comparable query lower bound still holds even when the algorithm is given a witness consisting of polynomially many random points $F$ from any one connected component.  

Next, we show that if there were a $\QCMA$ algorithm where the optimal witness depends only on the set of vertices $S$ in one of the connected components, by a counting argument, there must exist a combinatorial \emph{sunflower} of subsets $S$ that correspond to the same witness string. A \textit{sunflower}, in this context, is a set of subsets such that each subset contains a core $F \subset V$ and every vertex of $V \setminus F$ occurs in a small fraction of subsets. This implies that there exists a $\BQP$ algorithm which distinguishes $\YES$ instances corresponding to the sunflower from all $\NO$ instances. Next, we show using an adversary bound \cite{ambainis-lb}, a quantum query algorithm cannot distinguish the distribution of $\YES$ instances corresponding to the sunflower from the uniform distribution of $\YES$ instances such that the core $F$ is contained in a connected component (the ideal sunflower). 

This indistinguishability, along with the previous polynomial method based lower bound, proves that $\QCMA$ algorithm --- whose witness only depends on the vertices in the connected component --- for the expander distinguishing problem must make an exponential number of queries to the graph.

\paragraph{Removing the restriction over witnesses}

Our proof, thus far, has required the restriction that the witness only depends on the vertices in the connected component. In some sense, this argues that there is an oracle separation between $\QMA$ and $\QCMA$ if the prover is restricted to being "near-sighted": it cannot see the intricacies of the edge-structure of the graph, but can notice the separate connected components of the graph. If the near-sighted prover was capable of sending quantum states as witnesses, then she can still aid a verifier in deciding the expander distinguishing problem, whereas if she could only send classical witnesses, then she cannot aid a verifier.

It now remains to remove the restriction that the witness can only depend on the vertices in the connected component. We do this by introducing \emph{randomness} into the oracle, precisely designed to "blind" the prover to the local structure of the graph. In the standard oracle setting, the verifier and prover both get access to an oracle $x \in \bits^N$, and the prover provides either a quantum witness, $\ket{\xi(x)} \in \qubits{\poly(n)}$ or a classical witness, $\xi(x) \in \bits^{\poly(n)}$. The verifier then runs an efficient quantum algorithm $V^x$ which takes as input $\ket{\xi(x)}$ (or $\xi(x)$, respectively) and consists of quantum oracle gates applying the unitary transform defined as the linear extension of
\begin{equation}
    \ket{i} \mapsto (-1)^{x_i} \ket{i} \text{ for } i \in [N].
\end{equation}
We now extend modify this setup slightly. Instead of a single oracle $x$, we consider a distribution $\Bb$ over oracles. The prover constructs a quantum witness $\ket{\xi(\Bb)}$ (or a classical witness $\xi(\Bb)$, respectively) based on the distribution $\Bb$. The verifier then samples a classical oracle $x \leftarrow \Bb$ from the distribution, and then runs the verification procedure $V^x$ which takes as input $\ket{\xi(\Bb)}$ (or $\xi(\Bb)$, respectively) and applies quantum oracle gates corresponding to $x$. The success probability of the verifier is taken over the distribution $\Bb$ and the randomness in the verification procedure. 

\begin{figure}[h]
\begin{center}
\begin{tikzpicture}
\filldraw[fill=red!5] (0,2) -- (3,2) -- (3,7) -- (0,7) -- (0,2);
\filldraw[fill=blue!5] (6,2) -- (9,2) -- (9,7) -- (6,7) -- (6,2);
\node[text width=3cm,align=center] at (1.5,6.5) {\textbf{Prover}};
\node[text width=3cm,align=center] at (7.5,6.5) {\textbf{Verifier}};
\node[text width=3cm,align=center] at (4.5,6) {$\Bb$};
\draw[thick,->] (4.2,6) -- (3.1,6);
\draw[thick,->] (4.8,6) -- (5.9,6);
\node[text width=3cm,align=center] at (1.5,5.5) {$\ket{\xi} = \ket{\xi(\Bb)}$};
\node[text width=3cm,align=center] at (1.5,5) {(or $\xi = \xi(\Bb)$)};
\node[text width=3cm,align=center] at (4.5,4.5) {$\ket{\xi}$ (or $\xi$)};
\draw[thick,-] (3.7,4.5) -- (3.1,4.5);
\draw[thick,->] (5.3,4.5) -- (5.9,4.5);

\node[text width=3cm,align=center] at (7.5,5.25) {$x \leftarrow \Bb$};

\draw[thick,->] (7.5,4.9) -- (7.5,4.1);
\draw[thick,-] (6.1,4.5) -- (7.5,4.5);
\draw[thick,-] (6.1,6) -- (7.5,6);
\draw[thick,-] (2.9,6) -- (1.5,6);
\draw[thick,->] (1.5,6) -- (1.5,5.8);
\draw[thick,->] (7.5,6) -- (7.5,5.6);
\draw[thick,->] (1.5,4.5) -- (2.9,4.5);
\draw[thick,-] (1.5,4.5) -- (1.5,4.7);

\draw (6.5,2.5) -- (8.5,2.5) -- (8.5,4) -- (6.5,4) -- (6.5,2.5);
\node[text width=3cm,align=center] at (7.5,3.5) {$V^x(\ket{\xi})$};
\node[text width=3cm,align=center] at (7.5,3) {(or $V^x(\xi)$)};
\end{tikzpicture}
\end{center}
\caption{\it Cartoon of interaction between Prover and Verifier for a distribution over classical boolean functions.}
\end{figure}
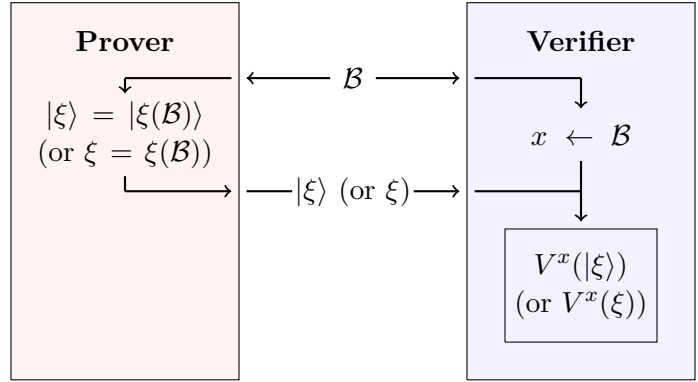

From our previous observations, graphs with the same connected component $S$ have the same ideal witness state (given in \eqref{eq:intro-ideal-classical-pf}). So, if the distribution $\Bb$ is supported on all graphs with the same connected component $S$, then the witness state from \eqref{eq:intro-ideal-classical-pf} suffices. Furthermore, in the case of the classical witness system, the witness can only depend on $S$ and the previously stated lower bound applies. 
This motivates the oracle problem of distinguishing distributions, marked either $\YES$ or $\NO$, over $2^n$ bit strings (or equivalently $n$-bit functions).

\subsection{Statement of the result}

\begin{theorem}
For every sufficiently large integer $n$ that is a multiple of 200, there exist distributions over $100$-regular $100$-colored graphs on $N = 2^{n}$ vertices labeled either $\YES$ or $\NO$ such that
\begin{itemize}
    \item 
    Each $\YES$ distribution is entirely supported on $\YES$ instances of the expander-distinguishing problem and, likewise, each $\NO$ distribution is entirely supported on $\NO$ instance of the expander-distinguishing problem. 
    \item There exists a $\poly(n)$ time  quantum algorithm $V_q$ taking a witness state $\ket{\xi}$ as input and making $O(1)$ queries to the quantum oracle such that
    \begin{enumerate}
        \item For every $\YES$ distribution $\Bb$, there exists a quantum witness $\ket{\xi} \in \qubits{n}$ such that
        \begin{equation}
            \Exp_{x \leftarrow \Bb} \Pr[V_q^{x}(\ket{\xi}) \text{ accepts}] \geq 1 - O(N^{-3}).
        \end{equation}
        \item For every $\NO$ distribution $\Bb$, for all quantum witnesses $\ket{\xi} \in \qubits{n}$,
        \begin{equation}
            \Exp_{x \leftarrow \Bb} \Pr[V_q^{x}(\ket{\xi}) \text{ accepts}] \leq 0.01.
        \end{equation}
    \end{enumerate}
    \item Any quantum algorithm $V_c$ accepting a classical witness of length $q(n)$ satisfying the following two criteria either requires $q(n)$ to be exponential or must make an exponential number of queries to the oracle. 
    \begin{enumerate}
        \item For every $\YES$ distribution $\Bb$, there exists a classical witness $\xi = \xi(\Bb) \in \bits^{q(n)}$
        \begin{equation}
            \Exp_{x \leftarrow \Bb} \Pr[V_c^{x}(\xi) \text{ accepts}] \geq 0.99.
        \end{equation}
        \item For every $\NO$ distribution $\Bb$, for all classical witnesses $\xi \in \bits^{q(n)}$,
        \begin{equation}
            \Exp_{x \leftarrow \Bb} \Pr[V_c^{x}(\xi) \text{ accepts}] \leq 0.01.
        \end{equation}
    \end{enumerate}
\end{itemize}
\label{thm:main}
\end{theorem}

\noindent Although our main theorem is formulated as a query lower bound, it can be converted to a separation between the relativized classes of $\QMA$ and $\QCMA$ via a standard diagonalization argument. Similarly, it was pointed out to us~\cite{honghaofu_2022} that it proves a separation between the relativized classes of $\BQP/\textsf{qpoly}$ and $\BQP/\textsf{poly}$, following the technique of \cite{ak-oracle}.

\subsection{Implications and future directions}

\noindent There are several future questions raised by this work that we find interesting:

\paragraph{Oracle and communication separations}
The most natural question is, of course, whether the oracle's randomness can be removed to obtain a separation in the standard model. We conjecture that our problem yields such a separation, but a new technique seems necessary to prove it. See Section \ref{sec:concluding-remarks} for more details on the technical barriers to derandomizing our construction.

Another natural question is to show a \emph{communication complexity} separation between $\QMA$ and $\QCMA$. This has been shown for one-way communication by Klauck and Podder~\cite{klauck2014two} but their problem does not yield a separation for two-way communication. Could our query separation be lifted to the communication world by use of the appropriate gadget?

The class $\QMA(2)$ is another relative of $\QMA$ which is perhaps even more enigmatic than $\QCMA$. In $\QMA(2)$, the witness state is promised to be an unentangled between the first and second half of the qubits. We do not even know of a quantum (unitary) oracle separation between $\QMA(2)$ and $\QMA$, nor do we have a natural candidate problem. Could we at least formulate such a candidate by considering "oracular" versions of $\QMA(2)$-complete problems, in analogy to what we do in this work for $\QCMA$.

\paragraph{Search-to-decision} In \cite{search-to-decision}, Irani, Natarajan, Nirkhe, Rao and Yuen studied the complexity of generating a witness to a $\QMA$ problem (equivalently, generating a ground state of a local Hamiltonian) when given \emph{oracle access} to a $\QMA$ oracle. This paradigm, called \emph{search-to-decision}, is commonplace in classical complexity theory (for example, $\P$, $\NP$, $\MA$, etc. all have search-to-decision reductions) and yet \cite{search-to-decision} gives evidence that $\QMA$ likely does not exhibit a search-to-decision reduction. They prove this by showing an oracle relative to which $\QMA$ search-to-decision reductions are provably impossible. The oracle used is identical to that of Aaronson and Kuperberg \cite{ak-oracle} to separate $\QMA$ and $\QCMA$. \cite{search-to-decision} acknowledge this noncoincidence and conjecture whether \emph{any} $\QMA$ and $\QCMA$ separating oracle yields a $\QMA$ search-to-decision impossibility result. Similar to the reasons for why the gold-standard of oracle separation between $\QMA$ and $\QCMA$ is a $n$-bit boolean function, the ideal oracle for proving $\QMA$ search-to-decision impossibility is also a $n$-bit boolean function. Does the oracle presented here also yield a search-to-decision impossibility?

\paragraph{Implications for Quantum PCPs}

The quantum PCP conjecture \cite{quant-ph/0210077} is one of the biggest open questions in quantum complexity theory. In a recent panel \cite{nirkhe-simons} on the quantum PCP conjecture and the NLTS theorem \cite{anshu2022nlts}, an interesting question was posed of whether $\MA$ or $\QCMA$ (lower or upper) bounds can be placed on the complexity of the promise-gapped local Hamiltonian problem. We recommend \cite{Nirkhe:EECS-2022-236} for an introduction to the subject. Because the oracle presented in this result corresponds to a sparse Hamiltonian with a problem of deciding if the second eigenvalue of the Hamiltonian is $1$ or $< 1 - \alpha/d = 1 - \Omega(1)$, one might wonder if this provides oracular evidence that quantum PCPs are at least $\QCMA$-hard. Unfortunately, to the best of our knowledge, this is not a reasonable conclusion. While we give evidence that the promise-gapped \emph{sparse} Hamiltonian problem is likely $\QCMA$-hard, the reduction from the sparse Hamiltonian problem to the local Hamiltonian problem does not imply that the promise-gapped local Hamiltonian problem is likely $\QCMA$-hard. The only algorithm known for checking a witness for the sparse Hamiltonian problem is Hamiltonian simulation on the witness which is not a local algorithm. 

\paragraph{Connections to Stoquastic Hamiltonians}
Since the oracles studied in this work correspond to the adjacency lists of graphs, they can be viewed as sparse access to a Hamiltonian $H$ which is the Laplacian of a graph (recall that if the adjacency matrix is $A$, then the Laplacian is $\II - A/d$). Such Hamiltonians have a special structure not present in general Hamiltonians: they are \emph{stoquastic}, meaning that the off-diagonal entries are nonpositive. The local Hamiltonian (LH) problem for stoquastic Hamiltonians is significantly easier than the general LH problem, and in some cases is even contained in $\MA$ as shown by Bravyi and Terhal~\cite{stoquastic-hams1}. It is worth noticing why this is not in tension with our result---in particular, why this does not imply that our oracle problem is contained in oracular $\MA$.

\begin{itemize}
    \item Crucially, the $\MA$-containment for stoquastic LH holds \emph{only} for the ground state: this is because of the Perron-Frobenius theorem, which implies that ground states of such Hamiltonians have nonnegative coefficients. However, in our case, we want the first excited state: the state of minimum energy for $H$ restricted to the subspace orthogonal to the uniform superposition. It was shown by \cite{osti_21408402} that all excited state energies are $\QMA$-hard to calculate for a stoquastic Hamiltonian.
    \item The $\MA$ containment also uses the locality of the Hamiltonian, which in turn imposes a strong structure on the adjacency matrix of the graph. The random graphs we consider will not have this structure. (While it was shown by \cite{stoquastic-hams2} showed an $\AM$ algorithm for calculating the ground energy \emph{stoquastic and sparse} Hamiltonians, again this does not apply to higher excited states.)
    \item At an intuitive level, in graph language, the LH problem for stoquastic Hamiltonians is to find a component of the graph where the average value of some \emph{potential function} (given by the diagonal entries of $H$) is minimized. An $\MA$ verifier can solve this by executing a random walk, given the right starting point by Merlin. In contrast, our problem is to determine whether the graph as a whole is connected---a global property which an $\MA$ verifier cannot determine.
\end{itemize}

\section{Organization of the paper}
The remainder of the paper is the proof of Theorem 1. The proof is divided into smaller components and these intermediate results are joined together in Section \ref{sec:putting-it-together}.
In Section \ref{sec:preliminaries}, we state some basic definitions and formally define the expander distinguishing problem.
In Section \ref{sec:pml-graph-construction}, we describe the distributions over graphs that constitute $\YES$ and $\NO$ instances.
In Section \ref{sec:qma-protocol}, we prove that there is an efficient $\QMA$ algorithm for the expander distinguishing problem. In particular, there is a single quantum witness that serves all the graphs in each of the $\YES$ distributions.
In Section \ref{sec:adversary-method}, we use the adversary method and counting arguments to prove that any $\QCMA$ algorithm for the expander distinguishing problem for the constructed distributions implies a $\BQP$ algorithm for distinguishing $\YES$ instances with a connected component corresponding to an ideal sunflower from a generic $\NO$ instance.
In Section \ref{sec:polynomial-method}, we argue using the polynomial method that such an algorithm is impossible without an exponential query complexity.
In Section \ref{sec:concluding-remarks}, we present some concluding remarks about our construction and its relation to other notions of computational complexity.
Appendices \ref{appendix:chernoff} and \ref{appendix:adversary-proofs} consist of omitted proofs.

\section{Preliminaries}
\label{sec:preliminaries}

\subsection{Notation and quantum information basics}

\label{subsec:notation}

We will assume that the reader is familiar with the basics of quantum computing and quantum information.
We will use $N \defeq 2^n$ throughout this paper and we will only consider graphs of $N$ vertices. The adjacency list of a $d$-regular $d$-colored graph on $N$ vertices takes $dnN$ bits to describe.
For any $m$, we abbreviate the set of integers $\{1,2,\ldots m\}$ as $[m]$. For a set $A\subseteq [N]$,
we will use $\ket{A}$ to denote the state $\frac{1}{\sqrt{A}} \sum_{j \in A}\ket{j}$, the subset state corresponding to $A$.
Unless otherwise, specified we assume $\norm{\cdot}$ is the Euclidean norm $\norm{\cdot}_2$ for a vector, and the spectral norm for a matrix, which is the largest singular value.

\subsection{Expander graphs}
\begin{definition}
A graph $G$ is a \emph{spectral $\alpha$-expander} (equiv. is $\alpha$-expanding) if the second highest eigenvalue $\lambda_2$ of the normalized adjacency matrix of $G$ satisfies $\lambda_2 \leq 1 - \alpha$. We say that a connected component $S$ of the graph is $\alpha$-expanding if the restricted graph to the vertices of $S$ is $\alpha$-expanding.
\label{def:expander}
\end{definition}
\begin{lemma}\label{lem:expander-mixing}
Let $G$ be a $d$-regular $\alpha$-expander. Consider the random walk that starts in any distribution over the vertices, and at each time step, stays in place with probability $1/2$, and moves along an edge of the graph with probability $1/2$. Then for any vertex $v$, after $\ell$ steps, the probability $\Pr[v]$ that the walk is in $v$ satisfies
\begin{equation} \abs{\Pr[v] - \frac{1}{N}} \leq \qty( 1- \frac{\alpha}{2})^\ell. \end{equation}
In particular, when $\ell = O(c \log N / \alpha)$ we can get the RHS to be $1/N^{c}$.
\end{lemma}
\begin{proof}

Let the  normalized adjacency matrix of $G$ be $A$, and let $A' = \frac{1}{2}(\II + A)$ be the transition matrix of the random walk.
If $G$ is a d-regular $\alpha$-expander then $(1/N) \mathbf{1}$ is the unique eigenvector of $A$ (and $A'$) with eigenvalue $1$. Moreover, since $\norm{A} \leq 1$, all eigenvalues of $A'$ are nonnegative. Since $G$ is an $\alpha$-expander, the second eigenvalue of $A$ is at most $1 - \alpha$, and thus the second eigenvalue of $A'$ is at most $1 - \alpha/2$.

Let $u$ be a vector representing a probability distribution over vertices of $G$ (i.e. $u \in \mathbb{R}_{+}^{N}$ with $\|u\|_1 = 1$), and let $\mathbf{1}$ be the $N$-dimensional all-ones vector. Then the statement we wish to prove is  equivalent to
\begin{equation} \norm{(A')^{\ell} u - \frac{1}{N} \mathbf{1} }_{\infty} \leq (1-\alpha)^{\ell}. \end{equation}
 Write $u = (1/N) \mathbf{1} + \delta$. By the condition that $1 = \| u\|_1 = \sum_i u_i = 1 + \sum_i \delta_i$, it holds that $\langle \delta_i, \mathbf{1} \rangle = 0$.  Then
\begin{xalign}
    \norm{ (A')^\ell u - \frac{1}{N} \mathbf{1} }_\infty &= \|(A')^\ell \delta\|_\infty \\
    &\leq \| (A')^\ell \delta \|_2 \\
    &\leq (1 - \alpha/2)^\ell \|\delta\|_2 \\
    &\leq (1 - \alpha/2)^\ell \|\delta\|_1 \\
    &\leq (1 - \alpha/2)^\ell.
\end{xalign}
Setting this quantity equal to $1/N^c$ and solving for $\ell$, we get
\begin{xalign}
    (1 - \alpha/2)^\ell &= N^{-c} \\
    \ell \log( 1 - \alpha/2) &= -c \log N \\
    \ell &= -2 \frac{\log N}{\log(1 - \alpha/2)} \\
        &\approx \frac{2 c\log N}{\alpha}.
\end{xalign}
\end{proof}

\subsection{Non-deterministic oracle problems}
\begin{definition}[Quantum oracle problems]
For a $n$-bit boolean function $\Oo$, we say an \emph{oracle} decision problem $\Ll^\Oo$ is in $\QMA^\Oo(\eps)$ if there exists a uniform family of quantum circuits $A^\Oo$ such that 
\begin{enumerate}
    \item For every $\YES$ instance $\Oo$, there exists a quantum state $\ket{\xi}$ of $\poly(n)$ qubits such that $A^\Oo(\ket{\xi})$ accepts with probability $\geq 1-\eps$.
    \item For every $\NO$ instance $\Oo$, for all quantum states $\ket{\xi}$ of $\poly(n)$ qubits, $A^\Oo(\ket{\xi})$ accepts with probability $\leq \eps$.
\end{enumerate}
$\QCMA^\Oo(c,s)$ is defined similarly, except the state $\ket{\xi}$ is promised to be classical. The classes $\QMA^\Oo$ and $\QCMA^\Oo$ are defined as $\QMA^\Oo(1/3)$ and $\QCMA^\Oo(1/3)$, respectively.
\end{definition}

We note that due to parallel repetition, $\QMA^\Oo(\eps = \half - 1/\poly(n)) = \QMA^\Oo = \QMA^\Oo(\eps = 2^{-\poly(n)})$. Likewise, for $\QCMA^\Oo$. This justifies removing the constant $\eps$ from the definition. We now define the same problem for oracles equaling distributions over $n$-bit boolean functions.

\begin{definition}[Random classical oracles]
A random oracle $\Rr$ is a distribution over classical oracles $\{\Oo\}$. We say an \emph{oracle} decision problem $\Ll^\Rr$ is in $\QMA^\Rr(\eps)$ if there exists a uniform family of quantum circuits $A^\Oo$ such that 
\begin{enumerate}
    \item For every $\YES$ instance $\Rr$, there exists a quantum state $\ket{\xi}$ of $\poly(n)$ qubits such that 
    \begin{equation}
        \Exp_{\Oo \in \Rr} \Pr \left[ A^\Oo(\ket{\xi}) \text{ accepts} \right] \geq 1-\eps.
    \end{equation}
    \item For every $\NO$ instance $\Oo$, for all quantum states $\ket{\xi}$ of $\poly(n)$ qubits, 
        \begin{equation}
        \Exp_{\Oo \in \Rr} \Pr \left[ A^\Oo(\ket{\xi}) \text{ accepts} \right] \leq \eps.
    \end{equation}
\end{enumerate}
$\QCMA^\Rr(c,s)$ is defined similarly, except the state $\ket{\xi}$ is promised to be classical. 
\end{definition}

Ideally, we would define the classes $\QMA^\Rr$ and $\QCMA^\Rr$ are defined as $\QMA^\Rr(1/3)$ and $\QCMA^\Rr(1/3)$, respectively. However, the parallel repetition argument for boolean function oracles cannot be extended to distributions over boolean functions. This is because the $\eps$ error that an algorithm is the expectation of the success probability of the algorithm over the distribution. It is possible that the algorithm runs on every instance in the distribution with error $\eps$ or it is possible that the algorithm succeeds with 0 error on a $1-\eps$ fraction of the distribution and fails on the remaining $\eps$ fraction. In the first case, the success of the algorithm can be improved with parallel repetition while it cannot in the second case\footnote{We note that this subtlety is overlooked in Fefferman and Kimmel \cite{fefferman-kimmel} but we believe that their result without parallel repetition is correct. Furthermore, the adversary bounds used in \cite{fefferman-kimmel} do not address this issue but can be rectified using the adversary bound stated in Theorem \ref{thm:ambainis-extension} which deals with \emph{average-case} distinguishing.}.

\subsection{Graph oracles}

\begin{definition}[Colored Graphs]
Given a $d$-colored $d$-regular graph $G = (V,E)$ on $N$ vertices, we say $G$ contains a triple $(j_1,j_2,\kappa) \in V^2 \times [d]$ if the edge $(j_1,j_2)$ exists in $G$ and is colored with color $\kappa$. 
\end{definition}

\begin{definition}[Adjacency graph oracles]
Let $G$ be a $d$-colored $d$-regular undirected graph. The graph $G = (V,E)$ can be described by an adjacency function
\begin{equation}
    \fn{G}{V \times [d]}{V}
\end{equation}
where the output of $(j, \kappa)$ returns the neighbor of $j$ along the edge colored with $\kappa$. Quantum access to the function $G$ is provided by the following \emph{oracle} unitary:
\begin{equation}
    \ket{j, \kappa, z} \overset{G}{\mapsto} \ket{j, c, z \oplus G(j,\kappa)}.
\end{equation}
We call the function $G$ the \emph{adjacency graph oracle} corresponding to $G$.
\end{definition}

\begin{definition}[Expander distinguishing problem]
The $(\alpha, \zeta)$-expander distinguishing problem is a promise oracle language where the input is an oracle $G$ for a $d$-colored $d$-regular undirected graph $G$ on $N$ vertices. The problem is to distinguish between the following two cases, promised that one holds:
\begin{itemize}
    \item $\YES$: the graph $G$ has a connected component $S$ of size at most $|S| \leq \zeta$. 
    \item $\NO$: the graph $G$ is an $\alpha$-expander.
\end{itemize}
\label{def:expander-distinguishing-problem}
\end{definition}
In this paper, we will think of $\alpha$ as a constant and $\zeta \sim N^{9/10}$.
To simplify notation, since the oracles considered in this result always correspond to graphs $G$, we express the algorithm as $\Aa^G$ rather than $\Aa^{\Oo}$.

\section{Random distributions over graphs with many connected components}

\label{sec:pml-graph-construction}

In this subsection, we describe distributions over graphs where the graphs with high probability consist of $\ell$ connected components. It should not be surprising that the distribution is almost identical to the distribution used by Ambainis, Childs, and Liu \cite{ambainis-childs-liu} in their proof that the expander distinguishing problem requires an exponential number of quantum queries for any quantum query algorithm in the absence of a proof. This is because we will reduce any $\QCMA$ algorithm to an efficient query algorithm for some expander distinguishing problem.

The lower bound in \cite{ambainis-childs-liu} is crucially a lower bound on the \emph{polynomial degree} of any polynomial that distinguishes two graph distributions. From there, it isn't too much to argue that these graph distributions are very close to $\YES$ and $\NO$ instances as prescribed in the expander distinguishing problem; therefore any algorithm solving the expander distinguishing problem must be able to distinguish these two graph distributions. Our first goal is to amplify the argument of \cite{ambainis-childs-liu} to a more restricted class of graphs.

\paragraph{\cite{ambainis-childs-liu} Graphs}

The goal of the construction is a distribution which depends on an integer $\ell$ and a subset $F \subset V$. The integer $\ell$ will roughly correspond to the number of connected components (henceforth denoted $C_1, \ldots, C_\ell$) in the graph and we insist that $F \subset C_1$. Every $v \in V \setminus F$ appears in each subset $C_i$ with equal probability of $1/\ell$.
The actual construction will be slightly more complicated than this but, morally, this is what we hope to achieve from the distribution.

\paragraph{Formal construction}

Let $N$ be an integer and for integer $M \geq N$, integer $\ell$ dividing $M$ and a subset $F \subset V$ define the distribution $P_{M,\ell}(F)$ over graphs on $N$ vertices as follows:

\begin{enumerate}
    \item Start by constructing a graph $G'$ on $M$ vertices: Partition $V'$ into $\ell$ equally sized sets of vertices $V_1, \ldots, V_\ell$. On each subset $V_k$, create a random \emph{colored} subgraph by randomly choosing $d$ perfect matchings (each with a different color $1, \ldots, d$) and taking their union.
    \item 
    {
    To construct the graph $G$ on $N$ vertices: We first choose an injective map $\iota: V \hookrightarrow V'$.  
    First, we pick a function $\fn{k}{V}{[\ell]}$. We pick $k$ as a uniformly random function conditioned on the fact that $k(j) = 1$ for each $j \in F$.
    }
    Let $\iota(v)$ be a random vertex from $V_{k(j)}$ \emph{without replacement} to satisfy injectivity. If all vertices from $V_{k(j)}$ have been selected with replacement, output the graph on $N$ vertices with no edges (i.e. abort).
    \item Induce a graph $G$ on $V$ from $G'$ and the map $\iota$ --- i.e. an edge $(j_1,j_2,\kappa) \in V^2 \times [d]$ exists if $(\iota(j_1), \iota(j_2), \kappa) \in {V'}^2 \times [d]$ is an edge.
    \item For a vertex $j$ and a color $\kappa$, if the previous induced edges did not introduce a $\kappa$-colored edge from $j$, then add edge $(j, j, \kappa)$.
    \item The distribution over graphs $G$ is henceforth called $P_{M, \ell}(F)$; when $F = \emptyset$, we write it as $P_{M,\ell}$.
\end{enumerate}

Notationally, for edges $e = (u,v,\kappa) \in G$, we will use $\iota(e) = (\iota(u), \iota(v), \kappa) \in G'$. Furthermore, we will extend $\iota$ naturally to subgraphs and subsets of vertices and edges.

\begin{remark}
For any $F$, $P_{M,1}(F) = P_{M,1}(\emptyset) \defeq P_{M,1}$.
\end{remark}

\subsection{Setting of constants}
\label{subsec:setting-of-constants}

The lower bounds we prove for the $\QCMA$ algorithm are by no means tight (up to constants). We make no attempt to perfect the choice of constants as our only goal is to prove an exponential lower bound on the size of the any quantum witness or the number of queries required to solve the expander distinguishing problem. For this reason, we pick the following constants:

\paragraph{Chosen constants} The degree of the graph $G'$ is set to be $d = 100$. We assume $\ell = N^{1/10}$, $\gamma = N^{-1/10}$ and $M = (1+\gamma)N$. 

\paragraph{Induced constants}
In Definition \ref{def:expander-distinguishing-problem}, we define an $(\alpha, \zeta)$-expander distinguishing problem. We will only consider $\alpha = 1/(2 \cdot 10^8)$ (which is a consequence of Lemma \ref{lem:good-expander-whp} and the chosen constants).
Notationally, we will use $z \defeq N/\ell = N^{9/10}$. We will use $\zeta = (1+\gamma)z = M/\ell$.

\paragraph{Conventions} Typically, we will assume (for the purposes of contradiction) that $\abs{F} \leq N^{1/100}$ but as that is a term we wish to bound, we explicitly state it each time. Anytime a set $S$ is described, it will be of size $\zeta$, but we will also state this.

\subsection{Concentration bounds for random distributions over graphs}

We will need the following concentration lemma about the generated distributions. The lemma proves that $P_{M,1}$ is approximately a $\YES$ instance and that $P_{M,\ell}(F)$ is approximately a $\NO$ instance. Overall, this lemma proves that any algorithm solving the expander distinguishing problem must do very well on identifying the distribution $P_{M,1}$ as a $\NO$ instance and identifying the distributions $P_{M,\ell}(F)$ as a $\YES$ instance. The proof of this lemma is provided in Appendix \ref{appendix:chernoff}.

\begin{restatable}[Adaptation of Lemma 16 of \cite{ambainis-childs-liu}]{lemma}{lemgoodexpanderwhp} \label{lem:good-expander-whp}

Assume $\abs{F} \leq N^{1/100}$. Then with  probability at least $\geq 1 - O(N^{-3})$, a graph drawn from distribution $P_{M,\ell}(F)$ consists of exactly $\ell$ connected components each $\alpha$-expanding and consisting of between $(1-\gamma)z$ and $(1+\gamma)z$ vertices. \\

\noindent Likewise, the probability that a graph drawn from the distribution $P_{M,1}$ is $\alpha$-expanding is $\geq 1 - O(N^{-3})$.
\end{restatable}

\noindent Note that being expanding necessarily implies connectivity. Note that when $F = \emptyset$ or $\ell = 1$, there are simpler proofs with tighter bounds but the bound proven here for the general statement is sufficient for our result. \\

\noindent The second concentration lemma that we will use is that $P_{M,\ell}$ is approximately equal to sampling a set $F$ of size $\leq N^{1/100}$ and then sampling a graph from $P_{M,\ell}(F)$. The proof of this lemma is also provided in Appendix \ref{appendix:chernoff}. 

\begin{restatable}{lemma}{lempmlclosetopmlf}
\label{lem:pml-close-to-pmlf} 
Let $m$ be $\leq N^{1/100}$. Let $\mathcal{D}_1$ be the distribution on pairs $(G,F)$ obtained by sampling $G \sim P_{M, \ell}$, choosing a uniformly random vertex $v \in G$, and then choosing $F$ to be a uniformly random subset of the connected component of $G$ containing $v$ of size $m$. Let $\mathcal{D}_2$ be the distribution on pairs $(G,F)$ obtained by first choosing $F$ to be a uniformly random subset of $V$ with size $m$, and then sampling $G \sim P_{M,\ell}(F)$. Then these distribution are close in statistical distance:
\begin{equation}\| \mathcal{D}_1 - \mathcal{D}_2\| \leq 3 N^{-9/200}.\end{equation}
\end{restatable}

\section{$\QMA$ protocol}
\label{sec:qma-protocol}

In this section we show that the expander distinguishing problem (over a fixed graph --- i.e., no distribution) can be solved with a polynomial number of queries (indeed, with just two queries) if a quantum witness is provided. Our algorithm has the added benefit of being time-efficient, so we have shown that this problem is contained in $\QMA^{G}$. In Section \ref{sec:putting-it-together}, we prove that there still exists a $\QMA$ protocol if we consider distribution oracles. 
\begin{lemma}
There is a $\QMA^{G}$ protocol $\Aa_\QMA$
that solves the $(\alpha, \zeta)$-expander distinguishing problem with the following properties:
\begin{enumerate}
    \item   \textbf{Query complexity:} the algorithm makes two queries to $G$.
    \item \textbf{Completeness:} In the $\YES$ case, there exists a witness state that the verifier accepts with certainty.
    \item \textbf{Soundness:} In the $\NO$ case, no witness state is accepted by probability greater than $1 - \alpha/4$.
    \item \textbf{Nice witnesses:} In the $\YES$ case, if $S \subsetneq V$ is a connected component of the graph $G$, then the state
    \begin{equation} \ket{S} = \frac{1}{\sqrt{|S|}} \sum_{v \in S} \ket{v} \end{equation}
    is accepted with probability at least $1 - \sqrt{|S|/N}$. 
    In particular, since there exists a connected component of size at most $\zeta$, there is a state of this form that is accepted with probability $1 - \sqrt{\zeta/N}$. 
\end{enumerate}
\label{lem:QMA-alg}
\end{lemma}
\begin{proof}

\newcommand{\wlk}{U_{\mathrm{walk}}}
First, we note the following fact: given access to the adjacency graph oracle for a $d$-regular graph, we can implement the unitary $\wlk$ defined by
\begin{equation} \wlk \ket{j,\kappa} \ket{\text{ancilla}} \mapsto \ket{G(j,\kappa), \kappa} \ket{\text{ancilla}},
\end{equation}
where $G(j,\kappa)$ is the $\kappa$-th neighbor of $j$ (which is guaranteed to exist by the $d$-regularity condition). To see this, prepare an ancilla in the $\ket{0}$ state, and apply
\begin{align}
    \ket{j,\kappa,0} &\overset{G}{\mapsto} \ket{j,\kappa, G(j,\kappa)} \overset{\text{swap registers}}{\mapsto} \ket{G(j,\kappa), \kappa, j } \overset{G}{\mapsto} \ket{G(j,\kappa), \kappa, 0}
\end{align}
Here, we have used the fact that for a validly colored graph, if $G(j_1,\kappa) = j_2$, then $G(j_2,\kappa) = j_1$. Moreover, using controlled queries to $G$, we can also implement the controlled version of $\wlk$. 

Let us also define the following states:
\begin{equation}
    \ket*{\overline{0}_V} := \frac{1}{\sqrt{N}} \sum_{v \in V} \ket{j}, \qquad \ket*{\overline{0}_d} := \frac{1}{\sqrt{d}} \sum_{\kappa=1}^{d} \ket{\kappa}.
\end{equation}

Let $\ket{\psi}$ be the witness state received from the prover. The verifier performs the following operation:
\begin{enumerate}

\item First, the verifier prepares an ancillas in the state $\ket{+}$ and $\frac{1}{\sqrt{d}} \sum_{\kappa =1}^{d} \ket{\kappa}$. The total state at this point is
    \begin{equation}
        \ket{+}_{\mathrm{control}} \otimes \ket{\psi}_{\mathrm{witness}}  \otimes \left(\frac{1}{\sqrt{d}} \sum_{\kappa=1}^{d} \ket{\kappa} \right)_{\mathrm{color}}.
    \end{equation}
    Now, the verifier applies $\wlk$ to the witness and color registers controlled on the control register.

\item Next, the verifier measures the control register in the $\{\ket{+}, \ket{-}\}$ basis, and the color register using the two-outcome measurement $\{M^C_0, M^C_1\}$ where $M^C_0 = \ket{\overline{0}_d}$. The verifier proceeds to the next step if and only if the control measurement yields $+$ and the color measurement yields $0$. Otherwise, it rejects.
\item Finally, it performs the two-outcome measurement $\{M^V_0, M^V_1\}$ on the witness register where $M^V_0 = \ketbra*{\overline{0}_V}$. If the outcome $0$ is obtained, it rejects. Otherwise, it accepts.
\end{enumerate}

The associated quantum circuit for this verifier is given in \Cref{fig:qma-verifier}.

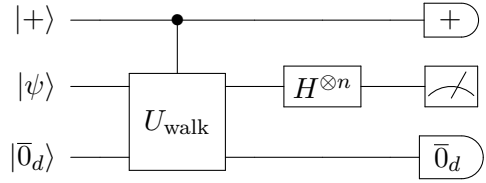
\begin{figure}[h]
$$
\Qcircuit @C=1em @R=1em {
 \lstick{\ket{+}} & \qw & \ctrl{1} & \qw & \qw & \qw & \measureD{+} \\
 \lstick{\ket{\psi}} & \qw & \multigate{1}{U_{\mathrm{walk}}} & \qw & \gate{H^{\otimes n}} & \qw & \meter \\
 \lstick{\ket*{\overline{0}_d}} & \qw & \ghost{U_{\mathrm{walk}}} & \qw & \qw & \qw & \measureD{\overline{0}_d}
}
$$
\caption{The $\QMA^{G}$ verifier.}
\label{fig:qma-verifier}
\end{figure}
\paragraph{Query complexity} From the description of the algorithm it is clear that only one query to $\wlk$, and thus two queries to $G$ are made.

\paragraph{Analysis}

To analyze the verification algorithm, let us write the witness as
\begin{equation}
    \ket{\psi} = \sum_j \alpha_j \ket{j}.
\end{equation}
Then the state after the controlled gate is
\begin{equation}
    \frac{1}{\sqrt{2}} \ket{0} \ket{\psi} \ket*{\overline{0}_d} +  \frac{1}{\sqrt{2}} \ket{1} \sum_j \sum_{\kappa \in [d]} \frac{\alpha_j}{\sqrt{\kappa}} \ket{G(j,\kappa)} \ket{i}.
\end{equation}
If we apply the projector of the third register onto $\ket*{\overline{0}_d}$, we get the resulting un-normalized state is
\begin{equation}
    \frac{1}{\sqrt{2}} \ket{0}\ket{\psi} + \frac{1}{\sqrt{2}} \ket{1} \sum_j \sum_{\kappa \in [d]} \frac{\alpha_j}{d} \ket{G(j,\kappa)} =  \frac{1}{\sqrt{2}} \ket{0}\ket{\psi} + \frac{1}{\sqrt{2}} \ket{1} A \ket{\psi} 
\end{equation}
where $A$ is the normalized adjacency matrix of the graph. 
The probability of measuring the control register as $\ket{+}$ on this un-normalized state is equal to to the probability that the control register and the color both yield accepting outcomes, and can be calculated to be
\begin{xalign}
    \Pr[+, \overline{0}_d] &= \left| (\bra{+} \otimes I) \left( \frac{1}{\sqrt{2}} \ket{0} \otimes \ket{\psi} + \frac{1}{\sqrt{2}}\ket{1}\otimes  A \ket{\psi} \right) \right|^2 \\
        &= \left| \frac{1}{2} \ket{\psi} + \frac{1}{2} A \ket{\psi} \right|^2 \\
        &= \frac{1}{4} \braket{\psi}{\psi} + \frac{1}{4} \bra{\psi} A^\dagger A \ket{\psi} + \frac{1}{2} \Re \bra{\psi} A \ket{\psi} \\
        &= \frac{1}{4} + \frac{1}{4} \bra{\psi} A^2 \ket{\psi} + \frac{1}{2} \bra{\psi} A \ket{\psi} \label{eq:test-score-exact} \\
        &\leq \frac{1}{2} + \frac{1}{2} \bra{\psi} A \ket{\psi}. \label{eq:test-score}
\end{xalign}
where in the penultimate line we used that $A$ is a real symmetric matrix, and in the last line we used that $A$ has operator norm at most $1$.

\paragraph{Completeness} In the $\YES$ case, let $S$ be a connected component and $T$ be its complement. Then the prover will send the state
\begin{equation}
    \ket{\xi_S} \defeq \sqrt{\frac{|T|}{N}} \ket{S} - \sqrt{\frac{|S|}{N}} \ket{T}. 
    \label{eq:intro-ideal-classical-pf}
\end{equation}
It is easy to check that $\ket{\xi_S}$ is a $+1$ eigenvector of $A$, and that it is orthogonal to $\ket*{\overline{0}_V}$. Therefore, by \eqref{eq:test-score-exact}, it is accepted by the verifier with probability $1$.

\paragraph{Soundness} Assume that for some $\NO$~case graph, the witness $\ket{\psi}$ is accepted with probability $1 - \gamma$ for $\gamma < \alpha/4$. Note the second eigenvalue $\lambda$ satisfies $|\lambda| \leq 1$, so $\delta = 1 - \lambda \leq 2$, and hence $1- \gamma > 1/2$. Since step 2 of the verifier must accept with at least this probability, by \eqref{eq:test-score}, \begin{equation} 1 - 2 \gamma \leq \ev{A}{\psi}.\end{equation}
 Since the $\NO$~case graph has a single connected component and a unique eigenvector of eigenvalue 1, $M_0^V$ is the projector onto the 1-eigenspace of $A$. Furthermore, since the graph is $\alpha$-expanding, every other eigenvector has eigenvalue at most $1 - \alpha$. Since $A$ commutes with $M_0^V$,
\begin{xalign}
    1 - 2 \gamma &\leq \ev{A}{\psi} \\
    &= \ev{M_0^V A M_0^V}{\psi} + \ev{(\II - M_0^V)A(\II - M_0^V)}{\psi} \\
    &\leq \norm{M_0^V \ket \psi}^2 + (1-\alpha)\left(1 - \norm{M_0^V \ket \psi}^2 \right) \\
        &= \alpha \norm{M_0^V \ket{\psi}}^2  + ( 1- \alpha)
\end{xalign}
Solving this equation tells us that
\begin{equation}
    \norm{M_0^V \ket \psi}^2 \geq 1 - \frac{2 \gamma}{\alpha}.     \label{eq:overlap-prob}
\end{equation}
Therefore, the measurement of the witness register in step 3 of the verifier will reject with probability $1 -  2\gamma/\alpha> 1/2$, causing the algorithm to reject with probability greater than $1/2$, a contradiction. Therefore, $\gamma \geq \alpha / 4$. \\

\paragraph{Nice witnesses} 
Grilo, Kerenedis and Sikora noted in \cite{subset-state-witnesses} that the witness for a $\QMA$ problem can always be assumed to be a subset state. We show that the same property holds for our oracular witnesses. The overlap between the ideal witness and $\ket{S}$ is
\begin{xalign}
\braket{\xi_S}{S} &= \sqrt{\frac{|T|}{N}}.
\end{xalign}
Therefore, the trace distance between these states satisfies
\begin{equation} \| \ketbra{\xi_S} - \ketbra{S} \|_1  \leq \sqrt{1 - \frac{|T|}{N}} = \sqrt{\frac{|S|}{N}}. \end{equation}
Observe that the entire operation of the verifier given $G$ can be modeled by a single two-outcome  measurement $\{M^G_{0}, M^{G}_{1}\},$ where $0$ corresponds to rejection and $1$ to acceptance. Therefore, by a standard trace distance fact \cite[Equation 9.22]{nielsen-chuang}, if $G$ is a $\YES$ instance, the probability that $\ket{S}$ is accepted is at least 
\begin{equation} \Pr[\text{$\ket{S}$ accepted}] \geq 1 - \sqrt{\frac{|S|}{N}}. \label{eq:prob-simple-witness-accepted}
\end{equation}

\end{proof}

\section{Adversary method}

\label{sec:adversary-method}

In this section, we use the adversary method of Ambainis to argue that any successful $\QCMA$ algorithm implies a $\BQP$ algorithm for distinguishing the distribution $P_{M,\ell}(F)$ and $P_{m,1}$ for $\abs{F} \leq N^{1/100}$. In Section \ref{subsec:improved-adversary} we state the adversary method result and in the following sections we prove the statement.

\subsection{Ambainis' proof of the adversary method}

\label{subsec:improved-adversary}
The adversary method of Ambainis \cite{ambainis-lb} is a convenient way of arguing lower bounds on the query complexity of oracular quantum algorithms. The adversary method lower bounds the complexity of any algorithm which (with high probability) computes $f(a)$ for a function $\fn{f}{\bits^N}{\bits}$. The quantum algorithm is allowed access to $a \in \bits^N$ by a oracle gate $O$ which applies linearly the transform $\ket{i} \mapsto (-1)^{a_i} \ket{i}$ for $i \in \bits^n$ (here $N = 2^n$). In doing so, the adversary method is a convenient way of producing $\BQP$ (query) lower-bounds.

To use it in our distributional setting, we make two modifications to the adversary bound. The first is to relax the notion of correctness. The lower bound of Ambainis is for a lower-bound for any algorithm which, for \emph{each} $a \in \bits^N$, outputs $f(a)$ correctly with probability $1-\eps$ for $\eps < \half$. We instead consider an \emph{average-case} notion of success in which
\begin{equation}
    \Exp_{a \in \bits^N} \Pr_{\Aa}\left[\Aa^a \neq f(a)\right] \leq \eps^2. 
\end{equation}
By Markov's inequality, this implies
\begin{equation}
    \Pr_{a \in \bits^N} \left[ \Pr_{\Aa}\left[\Aa^a \neq f(a)\right] \geq \eps \right] \leq \eps,
\end{equation}
or in other words, most $a$ are (with high probability) correctly identified. 

The second modification is to \emph{restrict} the set of locations that the algorithm is allowed to query the oracle. The reason for this is somewhat subtle. Essentially, the original lower bound of Ambainis was designed for decision problems with deterministic oracles, and relies on constructing a relation between two \emph{disjoint} sets of oracle instances, one consisting only of $\YES$ instances and the other only of $\NO$ instances. However, in our setting, we are interested in distinguishing two distributions over oracles that may have overlapping support. In order to define disjoint $\YES$ and $\NO$ sets of instances even when the distributions overlap, we add to each oracle string $a$ a set of flag bits $b$ that indicate which of the two distributions the string $a$ was sampled from. Naturally, any reasonable model cannot permit the algorithm to query the flag bits: otherwise, it would be easy to distinguish even two statistically close distributions with few queries.

More formally, we consider a generalization where the oracle string is a tuple $(a,b) \in \bits^N \times \bits^M$ and $\fn{f}{\bits^{N+M}}{\bits}$ but the algorithm can only query positions of $a$. 
In this model, with the average-case notion of success defined above, we obtain the following adversary lower bound for distributions:

\begin{restatable}{theorem}{thmambainisextension}
    Let $\fn{f}{\bits^{N+M}}{\bits}$ be a function and let $X,Y \subset \bits^N \times \bits^M$ be two subsets such that $X \subset \inv{f}(0)$ and $Y \subset \inv{f}(1)$. Let $R \subset X \times Y$ be a relation such that
    \begin{enumerate}
        \item For every $x \in X$, let $R_x \subset Y$ equal $R_x = \{y : (x,y) \in R\}$ such that $\underline{m} \leq \abs{R_x} \leq \overline{m}$.
        \item For every $y \in Y$, let $R_y \subset X$ equal $R_y = \{x : (x,y) \in R\}$ such that $\underline{m'} \leq \abs{R_x} \leq \overline{m'}$.
        \item For every $x = (a,b) \in X$ and $i \in [N]$, let $\ell_{x,i}$ be the number of $y = (c,d) \in Y$ such that $(x,y) \in R$ and $a_i \neq c_i$. Likewise, for every $y = (c,d) \in Y$ and $i \in [N]$, let $\ell_{y,i}$ be the number of $x = (a,b) \in X$ such that $(x,y) \in R$ and $a_i \neq c_i$. Let $\ell_{\max}$ be the maximum product $\ell_{x,i} \ell_{y,i}$ over $(x,y) \in R$ and $i \in [N]$ such that $a_i \neq c_i$.
    \end{enumerate}
    Then any quantum algorithm $\Aa$ which only queries the first $N$ bits of the oracle and computes $f$ such that
    \begin{equation}
        \Exp_{x = (a,b) \in X} \Pr_{\Aa}\left[\Aa^a \neq 0\right] \leq \eps^2 \quad \text{and} \quad \Exp_{y = (c,d) \in Y} \Pr_{\Aa}\left[\Aa^c \neq 1\right] \leq \eps^2  \label{eq:alg-assumption}
    \end{equation}
    uses
    \begin{equation}
        \geq \qty(1 - 2 \sqrt{\eps(1-\eps)}) \sqrt{\frac{\qty(\underline{m} - 2 \eps \overline{m})  \qty(\underline{m'} - 2 \eps \overline{m'})}{\ell_{\max}}} \quad \text{ queries}. \label{eq:thm-query-lb}
    \end{equation}
\label{thm:ambainis-extension}
\end{restatable}

\begin{restatable}{corollary}{corambainisextension}
Let $X$ and $Y$ be two subsets of $\bits^{N+M}$ satisfying the three conditions listed in Theorem~\ref{thm:ambainis-extension}. Then, any query algorithm $(1-\delta)$-distinguishing the uniform distributions over $X$ and $Y$, must use \eqref{eq:thm-query-lb} queries for $\eps = 2\delta$.
\label{cor:ambainis-extension}
\end{restatable}
The proofs of both statements are presented in Appendix \ref{appendix:adversary-proofs}.

\subsection{Setup from $\QCMA$ algorithm}

\label{subsec:setup-from-qcma-algorithm}

In this subsection, we show that if there is a $\QCMA$ algorithm for solving the \emph{expander distinguishing problem} then there exists a sunflower $\sunflower$ (defined below) of $\YES$ instances which correspond to the same optimal witness $\pf^\star$. If we hardcode $\pf^\star$ into the $\QCMA$ algorithm, we generate a quantum query algorithm that, with no access to a prover, accepts instances corresponding to $\sunflower$ and rejects all $\NO$ instances.

\begin{definition}[Sunflower]
A collection of subsets $\sunflower \subset {V \choose \zeta}$ is $(\mu, \zeta, t)$-\emph{sunflower} if there exists a subset $F \subset V$ with $\abs{F} \leq t$ satisfying the following two conditions:
\begin{enumerate}
    \item For all $S \in \sunflower$, $F \subseteq S$.
    \item For all $\displaystyle x \in \left(\bigcup_{S \in \sunflower} S\right) \setminus F$, the $\displaystyle \Pr_{S \in \sunflower}[x \in S] \leq \left(\frac{\zeta}{N}\right)^{1 - \mu}$.
\end{enumerate}
We call the set $F$ the core of the sunflower.
\end{definition}

\paragraph{$\YES$ instances corresponding to subsets}

For any graph $G$ and subset $S$ of size $\zeta$, define $G \triangleleft S$ if $G$ has a connected component $C_i \subseteq S$. Let $\S_{\triangleleft}$ be the set of $G$ such that $G \triangleleft S$. For each subset $S$ of size $\zeta$, define $B_S$ to be the restriction of the distribution $P_{M, \ell}$ to graphs in $\S_{\triangleleft}$. The intuition is that the witness $\ket{\xi_S}$ from \eqref{eq:intro-ideal-classical-pf} will be a good witness for $B_S$ since the connected components of $P_{M,\ell}$ are of a size concentrated around $z$.

There is a small complexity, which we address now, in that the distribution $P_{M, \ell}$ is not a uniform distribution over a set of graphs. To rectify this, we can always assume that the oracle corresponding to a graph $G$ sampled to $P_{M, \ell}$ consists of a queryable component corresponding to the adjacency list of $G$ and a non-queryable component corresponding to the random coins $r_G$ that were flipped in order to generate $G$ according to $P_{M, \ell}$. We will also define $B_S$ as the restriction of the extended oracle. Therefore, both $P_{M,\ell}$ and $B_S$ are uniform distributions over some support.

Lastly, the distributions $B_S$ are not exactly $\YES$ distributions since their support is not \emph{entirely} on $\YES$ graphs of the expander distinguishing problem. However, similar to Lemma \ref{lem:good-expander-whp}, we will show that $B_S$ is almost entirely supported on $\YES$ graphs. Therefore, it suffices to use $B_S$ as a \emph{proxy} for $\YES$ instances until the very end where we handle this subtlety.

\begin{corollary}
For every graph $G \in \S_{\triangleleft}$ such that $G$ has a connected component $C_i$ with $\abs{C_i} \geq (1-\gamma)z$, 
\begin{equation}\Pr[\Aa_\QMA^G(\ket{S}) = 1] \geq 1- 3\sqrt{\gamma}.
\end{equation}
\label{lem:accept-sub-optimal-pf}
\end{corollary}

\begin{proof}
Since, $\abs{C_i}/\abs{S} = \abs{C_i}/\zeta \geq (1-\gamma)/(1+\gamma) \geq 1 - 2\gamma$, then
\begin{equation}
    \norm{\ket{S} - \ket{C_i}} \leq \sqrt{2 \left(1-\sqrt{\frac{\abs{C_i}}{\abs{S}}}\right)} \leq \sqrt{2(1-\sqrt{1-2\gamma})} \leq 2\sqrt{\gamma}.
\end{equation}
By Lemma \ref{lem:QMA-alg}, $\ket{C_i}$ is accepted with probability $1 - \sqrt{(1-\gamma)z/N}$, then $\ket{S}$ is accepted with probability $1 - 3\sqrt{\gamma}$.
\end{proof}

\begin{corollary}
Let $M = (1+\gamma)N$ and $\gamma = N^{-1/10}$ and $\ell = N^{-1/10}$. For every $S$ of size $(1+\gamma)z$, the distribution $B_S$ is a $\YES$ instance and
\begin{equation}
    \Exp_{G \leftarrow B_S} \left[ \Pr \left[\Aa_\QMA^G(\ket{S}) = 1\right] \right] \geq 1 - 3\sqrt{\gamma} - O(N^{-3}).
\end{equation}
\label{cor:QMA-alg-completeness}
\end{corollary}

\begin{proof}
By Lemma \ref{lem:good-expander-whp}, with all but $O(N^{-3})$ probability, a graph $G$ drawn from the distribution $P_{M,\ell}$ consists of $\ell$ connected components each with size $\in [(1-\gamma)z, (1+\gamma)z]$. Then every connected component of the graph $G$ is contained in some connected component $S$ of size $(1+\gamma)z$. By the symmetry of the distribution $P_{M,\ell}$ under permutations, it follows that the distribution $\widetilde{P}_{M,\ell}$ formed by sampling a subset $S$ of size $(1+\gamma)z$ and then sampling a graph from the distribution $B_S$, is $O(N^{-3})$ close to the distribution $P_{M,\ell}$. Therefore, with probability all but $O(N^{-3})$, for any $S$, the distribution $B_S$ will consist of $\ell$ connected components each with size $\in [(1-\gamma)z, (1+\gamma)z]$. By the previous corollary, the algorithm with witness $\ket{S}$ accepts with probability $\geq 1 - 3\sqrt{\gamma}$. A union bound completes the proof.
\end{proof}

\paragraph{$\QCMA$ algorithm implies a quantum low-query algorithm for some sunflower $\sunflower$}

\begin{lemma}
For some $\eps > 0$, assume there exists a $k$-query non-deterministic quantum algorithm which accepts a $q$-length \emph{classical} witness and accepts every distribution $B_S$ for subset $S$ of size $\zeta$ with probability $1 - \eps$ and accepts any $\NO$ distribution $B_\NO$ with probability at most $\eps$.
Then for $\mu > 0$, there exists a $(\mu, \zeta, 2q/(\mu \log \ell) )$-sunflower $\sunflower$ and a $k$-query quantum algorithm that accepts every distribution $B_S$ for $S \in \sunflower$ and accepts any $\NO$ distribution $B_\NO$ with probability at most $\eps$.
\label{lem:qcma-implies-bqp-for-sunflower}
\end{lemma}

\begin{proof}
Assume such a non-deterministic algorithm $\Aa_\QCMA$ exists. Let the optimal witness (for algorithm $\Aa_\QCMA$) for oracle $B_S$ be $\pf(B_S)$; since the oracles are in bijection with subsets $S \in {V \choose \zeta}$, we can think of $\pf$ as function 
\begin{equation}
    \fn{\pf}{{V \choose \zeta}}{\bits^q}.
\end{equation} Formally this means that for every $S$, there exists a $k(n)$-query quantum algorithm $\Aa_\QCMA$ and a witness $\pf(S)$ such that
\begin{equation}
    \Exp_{G \leftarrow B_S} \left[ \Pr\left[ \Aa_\QCMA^{G}(\pf(S), 1^n) = 1\right]\right] \geq 1 - \eps.
\end{equation}
Let $\pf^\star \in \bits^q$ be the most popular witness (one associated with the largest number of subsets $S$) and let $\Sigma \defeq \inv{\pf}(\pf^\star)$. The size $\abs{\Sigma}$ is by a counting argument at least $2^{-q} {N \choose \zeta}$.
Notice that since $\abs{S} = \zeta$, then for a uniformly random $S \in {V \choose \zeta}$, $\Pr_S [j \in S] = \zeta/N$ for all $j \in V$. Ideally, if $S$ were instead sampled from $\Sigma$, we would like that $\Pr_{S \in \Sigma} [j \in S] \sim \zeta/N$ for all $j \in V$. Of course, this is too good to be true as $\Sigma$ could be the set of all $S$ such that $0^n$ is contained in $S$ (for example). Instead, we build a sunflower $\sunflower \subset \Sigma$ with the following greedy strategy inspired by \cite{fefferman-kimmel}. For $\sunflower = \Sigma$ (initially), whenever there exists a $j$ such that
\begin{equation}
    \Pr_{S \in \sunflower} [j \in S] \geq \left(\frac{\zeta}{N}\right)^{1-\mu},
\end{equation}
then we restrict $\sunflower \leftarrow \sunflower \cap \{S : j \in S\}$. By construction, after each restriction, the size of $\sunflower$ is at least its size before restriction multiplied by $(\zeta/N)^{1-\mu}$. If we continue this process and each time add $j$ to a set $F$, then
\begin{equation}
    \abs{\sunflower} \geq \left(\frac{\zeta}{N}\right)^{(1-\mu) \abs{F}} \abs{\Sigma} \geq \left(\frac{\zeta}{N}\right)^{(1-\mu) \abs{F}} \cdot 2^{-q(n)} \cdot \frac{N!}{\zeta!(N-\zeta)!}.
    \label{eq:sunflower-up}
\end{equation}
On the other hand, after selecting the set $F$, each element of $S \in \sunflower$ necessarily contains $F \subset S$ and therefore
\begin{equation}
    (S \setminus F) \subseteq {{V \setminus F} \choose {\zeta - \abs{F}}}.
\end{equation}
Therefore,
\begin{equation}
    \abs{\sunflower} \leq \frac{(N-\abs{F})!}{(\zeta-\abs{F})!(N-\zeta)!}.
    \label{eq:sunflower-down}
\end{equation}
Combining \eqref{eq:sunflower-up} and \eqref{eq:sunflower-down}, we get that
\begin{equation}
    \left(\frac{\zeta}{N}\right)^{(1-\mu) \abs{F}} \cdot 2^{-q(n)} \leq \frac{\zeta!}{N!} \cdot \frac{(N-\abs{F})!}{(\zeta-\abs{F})!} = \frac{\zeta}{N} \cdot \frac{\zeta-1}{N-1} \cdots \ldots \cdot \frac{\zeta - \abs{F} + 1}{N - \abs{F} + 1} \leq \qty(\frac{\zeta}{N})^{\abs{F}}. 
\end{equation}
Equivalently,
\begin{equation}
    \abs{F} \leq \frac{q(n)}{\mu \log (N/\zeta)} \leq \frac{2q(n)}{\mu \log \ell} . \label{eq:F-size-ub}
\end{equation}
The end result is that $\sunflower$ is $(\mu, \zeta, \frac{2 q(n)}{\mu \log \ell})$-sunflower.
Let $\Aa$ be the algorithm $\Aa_\QCMA$ with the witness $\pf^\star$ hard-coded into the algorithm. Since $\sunflower \subset \Sigma$, for every $S \in \sunflower$,
\begin{equation}
    \Exp_{G \leftarrow B_S} \left[ \Pr\left[ \Aa^{G}(1^n) = 1\right]\right] \geq 1 - \eps.
\end{equation}
And since the algorithm $\Aa_\QCMA$ accepts every $\NO$ distribution $B_\NO$ with probability at most $\eps$ irrespective of proof, then $\Aa$ also does them same.
\end{proof}
\subsection{Query lower bound for distinguishing sunflowers and fixed distributions}

Let $\isunflower \defeq {V \choose \zeta} \cap \{S : F \subseteq S\}$. This is the \emph{ideal sunflower} with a core of $F$. We will show by an adversary bound that the sunflower $\sunflower$ and the ideal sunflower $\isunflower$ are indistinguishable by quantum query algorithms with few queries.
Consider the distribution $H_\sunflower$ defined by sampling an $S \in \sunflower$ and then sampling a graph from $B_S$. Similarly, define the distribution $H_{\isunflower}$ but by first sampling an $S \in \isunflower$. We want to show that any quantum query algorithm requires exponentially many queries to distinguish $H_\sunflower$ and $H_{\isunflower}$.  The main result of this subsection is the following lemma.

\begin{lemma}
For $\delta < 1/4$, any quantum query algorithm $(1-\delta)$-distinguishing the distributions $H_\sunflower$ and $H_{\isunflower}$ where $\sunflower$ is a $(\mu, \zeta, t)$-sunflower and $F$ is the corresponding core requires \begin{equation}
    \geq \half \qty(1 - 2 \sqrt{2\delta(1-2\delta)})(1-4\delta) \cdot \sqrt{\qty(\frac{N}{\zeta})^{1-\mu}} \text{ queries}.
\end{equation}
\label{lem:distinguishing-sunflowers-and-fixed-distributions}
\end{lemma}

\subsubsection{A warmup lemma for distinguishing graphs}

The main challenge in proving \Cref{lem:distinguishing-sunflowers-and-fixed-distributions} is the complicated structure inherent in graphs. However, if we work instead directly with the \emph{sets} $S$, the problem is much simpler, and was already solved in~\cite[Lemma 11]{fefferman-kimmel}. They showed that given membership query access (equivalently, the indicator function for the set), it requires exponentially many quantum queries to distinguish a sample from $\sunflower$ from a sample from $\isunflower$.

We will work up to the result we wish to prove by gradually adding more structure to the objects being queried until we reach graphs. We will start by working with \emph{permutations} that map the set $S$ to a known set, and show that any algorithm with query access to the permutation and its inverse requires exponentially 
many queries.

To be precise, let $U = [\zeta]$. Let $\Pi_\sunflower$ be the set of all permutations and inverses $(\pi, \inv{\pi})$ such that $\pi(S) = U$ for some $S \in \sunflower$. Similarly, define $\Pi_{\isunflower}$. We shall abuse notation and also use $\Pi_\sunflower$ and $\Pi_{\isunflower}$ to refer to the uniform distributions over these sets of permutations. We first claim that no quantum query algorithm can distinguish the distributions $\Pi_\sunflower$ and $\Pi_{\isunflower}$ without an exponential number of queries. Note that the algorithm is allowed to query both the permutation and its inverse\footnote{This can be equivalently modeled by having a separate in-place oracle for the permutation and its inverse, or having a single "standard" oracle for the permutation.}.

\begin{lemma}
Any quantum query algorithm $(1-\delta)$-distinguishing the distributions $\Pi_\sunflower$ and $\Pi_{\isunflower}$ where $\sunflower$ is a $(\mu, \zeta, t)$-sunflower and $F$ is the corresponding core requires \begin{equation}
    \geq \qty(\frac{1}{2} - 2\sqrt{2\delta(1-2\delta)})(1-4\delta) \cdot \sqrt{\qty(\frac{N}{\zeta})^{1-\mu}} \text{ queries}.
\end{equation}
\label{lem:distinguishing-permutations}
\end{lemma}

\begin{proof}
We will use the adapted adversary bound (Theorem \ref{thm:ambainis-extension} and Corollary \ref{cor:ambainis-extension}) proved in Appendix \ref{appendix:adversary-proofs}. To do so we need to construct a relation $R \subset \Pi_{\sunflower} \times \Pi_{\isunflower}$. To build this relation, for every pair $(S_x, S_y) \in \sunflower \times \isunflower$, pick permutations $\chi_{xy}$ and $\psi_{xy}$ such that
\begin{enumerate}
    \item $\chi_{xy}(S_x) = \psi_{xy}(S_y) = U$.
    \item For all $j \in  (S_x \cap S_y) \cup \qty(V \setminus (S_x \cup S_y))$, $\chi_{xy}(j) = \psi_{xy}(j)$.
    \item For all $j_1 \in S_x \setminus S_y$, there exists a $j_2 \in S_y \setminus S_x$ such that $\chi_{xy}(j_1) = \psi_{xy}(j_2)$ and $\chi_{xy}(j_2) = \psi_{xy}(j_1)$.
\end{enumerate}
Such permutations are easy to find by picking a permutation $\chi_{xy}$ and then choosing the unique $\psi_{xy}$ that satisfies the constraints. Notice that every permutation mapping $S_x$ to $U$ can be expressed as $\tau \circ \chi_{xy}$ such that $\tau(U) = U$. Likewise, every permutation mapping $S_y$ to $U$ can be expressed as $\tau \circ \psi_{xy}$ for $\tau(U) = U$. Construct the relation $R$ by adding all pairs defined by the same $\tau$:
\begin{equation}
    R \defeq \left\{ \qty(\qty(\tau \circ \chi_{xy}, \inv{\chi_{xy}} \circ \inv{\tau}),\qty(\tau \circ \psi_{xy}, \inv{\psi_{xy}} \circ \inv{\tau})) ~:~ (S_x, S_y) \in \sunflower \times \isunflower, \tau(U) = U \right\}.
\end{equation}
Consider an element $(\pi, \inv{\pi}) \in \Pi_\sunflower$ such that $\pi(S_x) = U$. For any $S_y \in \isunflower$, write $\pi = \tau \circ \chi_{xy}$. Then $(\tau \circ \psi_{xy}, \inv{\psi_{xy}} \circ \inv{\tau}) \in \Pi_{\isunflower}$ forms a neighbor of $(\pi, \pi^{-1})$ along $R$. Therefore, the degree $m$ of every element of $\Pi_\sunflower$ is $\abs{\isunflower}$ and, analogously, the degree $m'$ of every element of $\Pi_{\isunflower}$ is $\abs{\sunflower}$.

Consider now any $\qty((\sigma_x, \inv{\sigma_x}), (\sigma_y, \inv{\sigma_y})) \in R$ such that $\sigma_x(S_x) = \sigma_y(S_y) = U$. To apply Theorem \ref{thm:ambainis-extension}, we need to calculate $\ell_{x,j}$ and $\ell_{y,j}$ where $\ell_{x,j}$ is the number of $y'$ such that $\qty((\sigma_x, \inv{\sigma_x}), (\sigma_{y'}, \inv{\sigma_{y'}})) \in R$ and $(\sigma_x, \inv{\sigma_x})$ and $(\sigma_y, \inv{\sigma_y})$ differ at index $j$. $\ell_{y,j}$ is defined analogously (see Theorem \ref{thm:ambainis-extension}). There are two cases to consider: (A) either an index $j$ such that $\sigma_x(j) \neq \sigma_y(j)$ or (B) an index $j$ such that $\inv{\sigma_x}(j) \neq \inv{\sigma_y}(j)$.
\begin{itemize}
    \item Case (A): By construction, in order for $\sigma_x(j) \neq \sigma_y(j)$, either (A1) $j \in S_x \setminus S_y$ or (A2) $j \in S_y \setminus S_x$.
    \begin{itemize}
        \item (A1): A simple upper bound for $\ell_{x,j}$ is $\abs{\isunflower}$. To bound $\ell_{y,j}$, notice that since $j \notin S_y$, then in order for $\sigma_{x'}(j) \neq \sigma_y(j)$, $j \in S_{x'} \setminus S_y$. The number of such $x'$ is equal to the number of sets $S_{x'}$ in $\sunflower$ that contain the point $\sigma_y(j)$.  Since $F \subset S_y$, it follows that the point $\sigma_y(j)$ is \emph{not} contained in $F$. Therefore, since $\sunflower$ is a sunflower, it holds that the number of $S_{x'} \in \sunflower$ satisfying the condition is at most
        \begin{equation} |\sunflower| \cdot \Pr_{S \in \sunflower}[\sigma_y(j) \in S] \leq \abs{\sunflower} \cdot \qty(\frac{\zeta}{N})^{1-\mu}.\end{equation}
          \item (A2): A simple upper bound for $\ell_{y,j}$ is $\abs{\sunflower}$. To bound $\ell_{x,j}$, notice that since $j \notin S_x$, then in order for $\sigma_{y'}(j) \neq \sigma_x(j)$, $j \in S_{y'} \setminus S_x$. Since $F \subset S_x$, then the number of such $y'$ is at most $\frac{\zeta-\abs{F}}{N-\abs{F}} \cdot \abs{\isunflower} \leq 2 \qty(\frac{\zeta}{N}) \abs{\isunflower}$.
    \end{itemize}
    \item Case (B): By construction, in order for $j' \defeq \inv{\sigma_x}(j) \neq \inv{\sigma_y}(j)$, either (B1) $j' \in S_x \setminus S_y$ or $j' \in S_y \setminus S_x$.
    \begin{itemize}
        \item (B1): Follows (A1) but for $j'$.
        \item (B2): Follows (A2) but for $j'$.
    \end{itemize}
\end{itemize}
Since this exhausts all cases,
\begin{equation}
    \ell_{\max} \defeq \max_{\substack{\qty((\sigma_x, \inv{\sigma_x}), (\sigma_y, \inv{\sigma_y})) \in R \\ \text{differing at } j}} \ell_{x,j} \ell_{y,j} \leq \qty(\frac{\zeta}{N})^{1-\mu} \abs{\sunflower}\abs{\isunflower}.
\end{equation}
Then by direct application of Corollary \ref{cor:ambainis-extension}, the algorithm must use
\begin{equation}
    \geq \qty(1 - 2\sqrt{2\delta(1-2\delta)})(1-4\delta) \cdot \sqrt{\qty(\frac{N}{\zeta})^{1-\mu}} \text{ queries}.
\end{equation}
\end{proof}

A short corollary of Lemma \ref{lem:distinguishing-permutations} is that there is a similar query lower bound for distinguishing distributions over graphs. Let $\Gg$ be a distribution over graphs with a connected component of $U = [\zeta]$. Let $\Gg_\sunflower$ be the distribution over graphs formed by sampling a permutation pair $(\pi, \inv{\pi})$ from $\Pi_\sunflower$, a graph $G$ from $\Gg$ and outputting the graph $\inv{\pi}(G)$. By construction, $\Gg_\sunflower$ is a distribution over graphs with a connected component of $S$ for $S \in \sunflower$. Likewise, define the distribution $\Gg_{\isunflower}$.

\begin{corollary}
For $\delta < 1/4$, any quantum query algorithm $(1-\delta)$-distinguishing the distributions $\Gg_\sunflower$ and $\Gg_{\isunflower}$ where $\sunflower$ is a $(\mu, \zeta, t)$-sunflower and $F$ is the corresponding core requires \begin{equation}
    \geq \half \qty(1 - 2\sqrt{2\delta(1-2\delta)})(1-4\delta) \cdot \sqrt{\qty(\frac{N}{\zeta})^{1-\mu}} \text{ queries}.
\end{equation}
\label{cor:distinguishing-graphs}
\end{corollary}

\begin{proof}
Any algorithm $\Aa$ for distinguishing $\Gg_\sunflower$ and $\Gg_{\isunflower}$, can be used as a subroutine in a (not necessarily time-efficient) algorithm $\Aa'$ for distinguishing $\Pi_\sunflower$ and $\Pi_{\isunflower}$ in twice as many queries.

To motivate the algorithm, observe that if $G$ is a random graph drawn from $\Gg$, then the graph $\pi^{-1}(G)$ is a random graph drawn from $\Gg_{\sunflower}$ or $\Gg_{\isunflower}$, depending on whether $\pi \in \Pi_{\sunflower}$ or $\Pi_{\isunflower}$. Moreover, a graph oracle query to the graph $\pi^{-1}(G)$ can be performed using two oracle queries to $\pi, \pi^{-1}$. 

Now we can specify the algorithm $\Aa'$: first, it samples a graph $G$ from $\Gg$---this step is not time-efficient, but it makes no oracle queries. Next, it runs $\Aa$ on the graph $\pi^{-1}(G)$, simulating oracle queries to this graph as described above. It answers according to the outcome of $\Aa$. If $\Aa$ successfully distinguishes $\Gg_{\sunflower}$ and $\Gg_{\isunflower}$, then $\Aa'$ distinguishes $\Pi_{\sunflower}$ and $\Pi_{\isunflower}$ with the same probability, and using twice as many oracle queries, as claimed.
Thus, by \Cref{lem:distinguishing-permutations}, we obtain the claimed query bound for $\Aa$.
\end{proof}

\subsubsection{Improving to more general permutations}

While \Cref{lem:distinguishing-permutations} and \Cref{cor:distinguishing-graphs} are simple enough to prove, they are insufficient at proving indistinguishability for the graph distributions $H_\sunflower$ and $H_{\isunflower}$ defined at the start of this section. This is because, unlike the distribution $\Gg_\sunflower$, the distribution $H_\sunflower$ cannot be defined in terms of independently sampling a graph $G$ and a set $S$. For one, the sizes of the connected components in $H_\sunflower$ do not exactly equal $z$; instead, the concentrate tightly around $z$.
It was precisely the independence of the graphs and sets that made Corollary \ref{cor:distinguishing-graphs} easy to prove. 

To fix the argument, we prove the following variations of Lemma \ref{lem:distinguishing-permutations} and Corollary \ref{cor:distinguishing-graphs}.
For a sunflower $\sunflower$ with core $F$ and any $k$ such that $\abs{F} \leq k \leq \zeta$, let $\Pi_\sunflower^{(k)}$ be the distribution formed by the following procedure:
\begin{enumerate}
    \item Sample a set $S$ from $\sunflower$.
    \item Sample uniformly randomly a subset $C \subset S$ of size $k$.
    \item Sample uniformly randomly a permutation $\fn{\pi}{V}{V}$ such that $\pi(C) = [k]$.
    \item Output $(\pi, \inv{\pi})$.
\end{enumerate}
Define the distribution $\Pi_{\isunflower}^{(k)}$ similarly where we change the first step to sampling from $\isunflower$.
\begin{lemma}
Any quantum query algorithm $(1-\delta)$-distinguishing the distributions $\Pi_\sunflower^{(k)}$ and $\Pi_{\isunflower}^{(k)}$ where $\sunflower$ is a $(\mu, \zeta, t)$-sunflower and $F$ is the corresponding core requires
\begin{equation}
    \geq \qty(1 - 2\sqrt{2\delta(1-2\delta)})(1-4\delta) \cdot \sqrt{\qty(\frac{N}{\zeta})^{1-\mu}} \text{ queries}.
\end{equation}
\label{lem:distinguishing-permutations-k}
\end{lemma}

\begin{proof}
This proof is equivalent to that of Lemma \ref{lem:distinguishing-permutations} except we use $U = [k]$. Note, the listed bound has no dependence on $k$; this is because $k \leq \zeta$ and we express here the weaker bound with $\zeta$.
\end{proof}
Likewise, a short corollary of Lemma \ref{lem:distinguishing-permutations-k} is the following. Construct the distribution $\Gg_\sunflower^{(k)}$ by the following procedure:
\begin{enumerate}
    \item Sample a graph $G$ from the restriction of the distribution $P_{M,\ell}$ to graphs with a connected component of exactly $[k]$.
    \item Sample a permutation $(\pi, \inv{\pi})$ from $\Pi_\sunflower^{(k)}$.
    \item Output $(\inv{\pi}(G), r_G)$ where $r_G$ is the random coin flips that would have generated $G$ when sampling according to $P_{M,\ell}$. The oracle will be divided into a queryable component of $(\inv{\pi}(G))$ and a un-queryable component of $r_G$.
\end{enumerate}

\begin{corollary}
For $\delta < 1/4$, any quantum query algorithm $(1-\delta)$-distinguishing the distributions $\Gg_\sunflower^{(k)}$ and $\Gg_{\isunflower}^{(k)}$ where $\sunflower$ is a $(\mu, \zeta, t)$-sunflower and $F$ is the corresponding core requires \begin{equation}
    \geq \half \qty(1 - 2\sqrt{2\delta(1-2\delta)})(1-4\delta) \cdot \sqrt{\qty(\frac{N}{\zeta})^{1-\mu}} \text{ queries}.
\end{equation}
\label{cor:distinguishing-graphs-k}
\end{corollary}
\begin{proof}
The corollary follows from \Cref{lem:distinguishing-permutations-k} via a reduction from permutations to graphs exactly as in the proof of \Cref{cor:distinguishing-graphs} from \Cref{lem:distinguishing-permutations}.
\end{proof}

\subsubsection{Completing the proof}

\begin{proof}[Proof of \Cref{lem:distinguishing-sunflowers-and-fixed-distributions}] Notice that for any $k \neq k'$, the support of $\Gg_\sunflower^{(k)}$ is disjoint from the support of $\Gg_{\sunflower}^{(k')}$, and likewise for $\Gg_{\isunflower}^{(k)}$ and $\Gg_{\isunflower}^{(k')}$. Let us again abuse notation and use $\Gg_\sunflower^{(k)}$ to denote the support of the corresponding distribution. For each $k$, the lower bound from \Cref{cor:distinguishing-graphs-k} is shown via an adversary bound with a relation $R_k$, and parameters $m, m', \ell_{\max}$, and moreover these parameters are the same for all $k$. Thus, we may construct a relation $R$ between $\bigcup_{k} \Gg^{(k)}_{\sunflower}$ and $\bigcup_{k} \Gg^{(k)}_{\isunflower}$ by simply taking the union $R = \bigcup_k R_k$. This relation maintains the same parameters $m, m', \ell_{\max}$ due to the disjointness of supports for different $k$. 
Lastly, notice that $\bigcup_k \Gg_\sunflower^{(k)}$ is equal to the support of $H_\sunflower$ as described in the statement of Lemma \ref{lem:distinguishing-sunflowers-and-fixed-distributions}. Likewise, for $H_{\isunflower}$. Since $H_\sunflower$ and $H_{\isunflower}$ are uniform distributions over their support, by Corollary \ref{cor:ambainis-extension} using the relation $R$ that we have constructed, the distributions are indistinguishable without the stated number of queries.

\end{proof}

\subsection{Statistical indistinguishability between random distributions}

The final step of this section is to show that no algorithm can distinguish the distributions $H_{\isunflower}$ and $P_{M,\ell}(F)$ with more than a negligible probability.
This will be because these distributions are statistically close and this can be proven by a Chernoff tail bound.

\begin{lemma}
The statistical distance between $H_{\isunflower}$ and $P_{M,\ell}(F)$ is $O(N^{-3})$.
\label{lem:chernoff-tail-bound}
\end{lemma}

\begin{proof}
Notice that the distribution $H_{\isunflower}$ is equivalent to sampling a graph from $P_{M,\ell}(F)$ conditioned on consisting of $\ell$ connected components each with size $\in [(1-\gamma)z, (1+\gamma)z]$.  By Lemma \ref{lem:good-expander-whp}, with all but $O(N^{-3})$ probability, a graph from $P_{M,\ell}(F)$ satisfies this condition. Therefore, the statistical distance between these distributions is bounded by $O(N^{-3})$.
\end{proof}

\section{Polynomial method lower bound}

\label{sec:polynomial-method}

In this section, we prove that any quantum query algorithm cannot distinguish the graph distributions $P_{M,1}$ and $P_{M,\ell}(F)$. When $F = \emptyset$, this is equivalent to the problem studied by \cite{ambainis-childs-liu} in their quantum query lower bound:

\begin{theorem}[Restatement of Theorem 2 of \cite{ambainis-childs-liu}]
\label{thm:acl-restatement}
For any sufficiently small constant $\eps_1 > 0$, any \emph{deterministic} quantum query algorithm $\Aa$ distinguishing the distributions $P_{M,1}$ and $P_{M,\ell}$ for any $1 < \ell < N^{1/4}$ by probability $\eps_1$. I.e.
\begin{equation}
    \Exp_{G \leftarrow P_{M,1}} \left[ \Pr_{\Aa}\left[\Aa^G = 1 \right] \right] - \Exp_{G \leftarrow P_{M,\ell}} \left[ \Pr_{\Aa}\left[\Aa^G = 1 \right] \right] \geq \eps_1
\end{equation}
must make at least $\Omega(N^{1/4}/ \log N)$ queries. Here the $\Omega$ notation hides a dependence on $\eps_1$.
\end{theorem}

The proof used in that result is very technical and builds on the polynomial method. Fortunately, we can show our query lower bound via a \emph{reduction} to the \cite{ambainis-childs-liu} result. The reduction requires taking a short walk which mixes well by the expander mixing lemma.

\begin{lemma}
Suppose there exists some $F_0$ and a $q_1$-query quantum algorithm that $\eps_1$-distinguishes the distributions $P_{M,1} = P_{M,1}(F_0)$  and $P_{M,\ell}(F_0)$ for $\ell > 1$. Then there exists a $q_2$-query quantum algorithm that $\eps_2$-distinguishes the distributions $P_{M,1}$ and $P_{M,\ell}$ with $q_2 = q_1 + O(N^{3/100})$ and $\eps_2 = \eps_1 - O(N^{-9/200})$. 
\label{lem:remove-F}
\end{lemma}

\noindent Intuitively, what this lemma says is that the set of points $F_0$ (which are in the same connected component) is not a helpful witness. Concretely, such a witness is negligibly more helpful than no witness at all. This is because, in the case of $P_{M,1}$ or $P_{M,\ell}$, the connected components are expanding and therefore the verifier can easily select a random subset of the points from a single connected component without any assistance from the prover. This can be shown via an application of the expander mixing lemma. Therefore, if a query algorithm exists for distinguishing $P_{M,1}$ and $P_{M,\ell}(F)$, it can be used as a subroutine for distinguishing $P_{M,1}$ and $P_{M,\ell}$ without any witness.

Furthermore, due to Ambainis, Childs, and Liu \cite{ambainis-childs-liu}, we know Theorem \ref{thm:acl-restatement} --- i.e. that distinguishing the distributions without witnesses has a query lower bound. Therefore, the problem has a query lower bound even when a set of points $F$ from a connected component are provided:

\begin{corollary}
For any $F_0$ with $|F_0| \leq N^{1/100}$, any sufficiently small constant $\eps_1$, and any $\ell$ with $1 < \ell < N^{1/4}$, any quantum query algorithm to $\eps_1$-distinguish $P_{M,1}$ and $P_{M,\ell}(F_0)$  must make $\Omega(N^{1/4}/ \log N)$ queries.
\label{cor:poly-method-lb}
\end{corollary}
\begin{proof}[Proof of \Cref{cor:poly-method-lb}]
Suppose an algorithm making $q = o(N^{1/4}/ \log N)$ queries existed. Then by \Cref{lem:remove-F} there exists an algorithm making $q' = q + O(N^{3/100}) = o(N^{1/4} / \log N)$ queries that distinguishes between $P_{M,1}$ and $P_{M,\ell}$ as well. However, this is impossible by \Cref{thm:acl-restatement}.
\end{proof}

\noindent The remainder of this section is the proof of \Cref{lem:remove-F}. \\

\noindent Let $\Aa_0$ be the hypothesized algorithm making $q_1$ queries to $\eps_1$-distinguish $P_{M,1}$ and $P_{M,\ell}(F_0)$. We first claim that for any $F$ with $|F| = |F_0|$, there exists an algorithm $\Aa_1$ that, given as classical input a list of all the vertices in $F$, and as oracle input an oracle $G$ where $G$ is a sample from either $P_{M,1}$ or $P_{M,\ell}(F)$, can $\eps_1$-distinguish between these two cases using $q_1$ queries to $G$. The algorithm $\Aa_1$ is as follows:
\begin{enumerate}
    \item Given $F$, compute a permutation $\pi$ on $V$ that maps $F$ to $F_0$. (This step is not efficient in terms of runtime, but makes no queries to the oracle $G$.)
    \item Run $\Aa_0$ with every query to $G$ replaced by a query to $\pi(G)$. Return the answer given by $\Aa_0$.
\end{enumerate}
The correctness of the algorithm follows from the fact that $\pi$ maps the distribution $P_{M,\ell}(F)$ exactly to $P_{M, \ell}(F_0)$. Therefore, for all $F$ such that $\abs{F} = \abs{F_0}$,
\begin{equation}
    \Exp_{G \leftarrow P_{M,\ell}(F)} \left[ \Pr_{\Aa_1}[\Aa_1(F,G) = 1]\right] - \Exp_{G \leftarrow P_{M,1}} \left[ \Pr_{\Aa_1}[\Aa_1(F,G) = 1]\right] \geq \eps_1.
\end{equation}
As this holds for all such $F$,
\begin{equation}
\Exp_{F} \Exp_{G \leftarrow P_{M,\ell}(F)} \left[ \Pr_{\Aa_1}[\Aa_1(F,G) = 1]\right] - \Exp_F \Exp_{G \leftarrow P_{M,1}} \left[ \Pr_{\Aa_1}[\Aa_1(F,G) = 1]\right] \geq \eps_1. \label{eq:success-of-a1}
\end{equation}

Next, we will show that the input of $F$ can be removed from the algorithm: given just access to $G$, it is possible to compute a suitable $F$ without making too many queries to the oracle. Specifically, we define the algorithm $\mathcal{A}_2$ to distinguish between $P_{M,1}$ and $P_{M,\ell}$ given only oracle access to $G$.
\begin{enumerate}
    \item For a choice of $t$ to be defined later, construct a set $F_1$ by starting at a random vertex $v_0$ and taking a $100t \cdot N^{1/100}$-step random walk along the graph as described in \Cref{lem:expander-mixing}. If $\abs{F_1} \geq \abs{F}$, pick the first $\abs{F}$ points from $F_1$ as the set $F'$. If not, output 0 (i.e. abort).
    \item Run $\mathcal{A}_1$ on input $F'$ with oracle access to $G$.
\end{enumerate}

We will argue that for an appropriately chosen $t$, this algorithm achieves the success probability and query complexity claimed in the theorem. To do so, we will argue in two stages.
\begin{enumerate}
    \item First, we argue that the distribution of $F'$ chosen by random walk is very close to $F'$ chosen uniformly at random from subsets of a connected component of $G$. This analysis uses the expander mixing lemma.
    \item Second, we argue that the distribution over pairs $(G,F')$ obtained after the first step of $\Aa_2$ is statistically indistinguishable from the distribution over pairs $(G,F)$ sampled by first choosing a uniformly random $F \subseteq V$ and then choosing a random $G \leftarrow P_{M, \ell}(F)$. This will make use of \Cref{lem:pml-close-to-pmlf}, shown in \Cref{appendix:chernoff}. By \eqref{eq:success-of-a1}, the algorithm $\Aa_1$ can $\eps_1$-distinguish inputs distributed in this manner, and thus the second step of $\Aa_2$ can $\eps_2$-distinguish inputs of $P_{M,1}$ and $P_{M,\ell}$ for $\eps_2$ just slightly smaller than $\eps_1$.

\end{enumerate}

In our analysis, we will denote probabilities over the distribution of $(G, F')$ generated by $\Aa_2$ by $\displaystyle \Pr_{\Aa_2}[\cdot]$ and probabilities over the distribution of $(G, F)$ obtained by first sampling $F \subseteq V$, and then sampling $G \leftarrow P_{M, \ell}(F)$ by $\displaystyle  \Pr_{\ftheng}[\cdot]$. The notation $\displaystyle \Pr_\unif[\cdot]$ denotes the distribution over $F$ obtained by first picking a uniformly random vertex $v$ in $G$, and then picking $F$ to be a uniformly random subset of the connected component of $G$ containing $v$ with size $|F_0|$.

\subsection{From random walk sampling to uniform sampling} 

Henceforth, define \emph{expander walk sampling} as the sampling procedure of selecting a uniformly random vertex as the initial vertex $v_1$, then subsequently taking $t$ steps of a lazy random walk (as defined in the expander mixing lemma, \Cref{lem:expander-mixing}) to choose $v_2$, and so forth. In this case, the graph and the integer $t$ will be clear from context.

We start by showing a sequence of claims that establish that if the expander walk sampling procedure for generating $F'$ starts in a connected component $C$ of $G$ with size $|C| = K$ and expansion $\alpha$, then the distribution over sets $F'$ generated by the random walk is close to uniformly sampling points from $C$. Our main result here will be \Cref{claim:walk-close-to-unif-sets}.

\begin{claim}\label{claim:walk-close-to-unif-points} 
Let $\delta = ( 1- \alpha/2)^t$ and let $r$ be a natural number with $r K\delta <1$. The for any sequence of $r$ vertices $v_1, \dots, v_r$, the probability $\displaystyle \Pr_{\unif}[\cdot]$ that this sequence was obtained by iid random sampling and the probability $\displaystyle\Pr_{\mathrm{walk}}[\cdot]$ that it was obtained by expander walk sampling differ by
\begin{equation} \abs{\Pr_{\unif}[v_1, \dots, v_r] - \Pr_{\mathrm{walk}}[v_1, \dots, v_r]} \leq \qty(\frac{1}{K})^r \cdot \qty( r K \delta+ (rK\delta)^2 \frac{1}{1 - r K\delta }) . \end{equation}
\end{claim}
\begin{proof}
The proof is by direct calculation and application of the expander mixing lemma. First we show one side of the bound.
\begin{xalign}
\Pr_{\unif}[v_1, \dots, v_r] &= \qty(\frac{1}{K})^r \\
\Pr_{\mathrm{walk}}[v_1, \dots, v_r] &\leq \qty(\frac{1}{K} + \delta)^r \\
&= \qty(\frac{1}{K})^r \cdot \qty(1 + K \delta)^r \label{eq:walk-seq-prob}
\end{xalign}
By hypothesis $K \delta < 1$. Using this, we have the estimate
\begin{xalign}
    (1 + K\delta)^r &= 1 + r K\delta + \sum_{j=2}^{r} \binom{r}{j} (K \delta)^{j} \\
    &\leq 1 + r  K\delta +  \sum_{j=2}^{r} (r K \delta)^j   \\
    &\leq 1 + r K \delta + (r K \delta)^2 \frac{1}{1 - (rK\delta)}
\end{xalign}
Substituting this bound into \Cref{eq:walk-seq-prob}, we obtain
\begin{align}
    \Pr_{\mathrm{walk}}[v_1, \dots, v_r] - \Pr_{\unif}[v_1, \dots, v_r]  &\leq \qty(\frac{1}{K})^r \cdot \qty(r  K\delta + (r K \delta)^2 \frac{1}{1 - rK\delta}).
\end{align}
Now for the other side. Again, applying the expander mixing lemma,
\begin{xalign}
    \Pr_{\mathrm{walk}}[v_1, \dots, v_r] &\geq \qty(\frac{1}{K} - \delta)^r \\
    &= \qty(\frac{1}{K})^r \qty(1 - K \delta)^r \\
    &\geq \qty(\frac{1}{K})^r \qty(1 - r K \delta),
\end{xalign}
where in the last step we have used Bernoulli's inequality and the assumption that $K \delta < 1$.
\end{proof}

The following claim will be used to bound the probability that the expander walk sampling procedure aborts, by instead bounding the probability that iid sampling fails to generate enough distinct points.
\begin{claim}\label{claim:unif-doesnt-abort}
The probability that $ T \geq 100 \abs{F}$ iid samples from $C$ contain fewer than $|F|$ distinct vertices is at most $\exp(-T/16)$. 
\end{claim}
\begin{proof}
Let $X_v$ be the event that vertex $v \in C$ is sampled. It is clear that
\begin{xalign}
    \Exp[X_v] &= 1 - \qty(1 - \frac{1}{K})^T \\
    &= \frac{T}{K} -\sum_{j=2}^{T} \binom{T}{j} \cdot \qty(\frac{-1}{K})^j \\
    &\geq \frac{T}{K} - \sum_{j=2}^{T} \qty(\frac{T}{K})^j \\
    &\geq \frac{T}{K} - \qty(\frac{T}{K})^2 \frac{1}{1 - T/K}.
\end{xalign}

In our setting $K \geq (1 - \gamma)z \geq N^{0.5}$ and $T = 100 |F| = 100 N^{1/100}$, so for sufficiently large $N$ we have
\begin{equation}
    \frac{T}{K} \geq \Exp[X_v] \geq \frac{T}{K} - \frac{2 \cdot 10^4}{K \cdot N^{0.48}} \geq \frac{T-1}{K}.
\end{equation}

So the expected number of vertices that are sampled is
\begin{xalign}
   \mu \defeq \Exp[N_{\text{sampled}}] &= \sum_{v} \Exp[X_v] \\
    &\geq T- 1.
\end{xalign}

Moreover, each even $X_v$ is independent, so the number of sampled vertices concentrates well around its mean. By a Chernoff bound we have
\begin{align}
    \Pr[ N_{\text{sampled}} \leq (1 - \eps) \mu] < \exp\qty(-\frac{\eps^2 \mu}{2}).
\end{align}
Setting $\eps = 0.5$ we have
\begin{align}
    \Pr[N_{\text{sampled}} \leq 50 |F| -1 ] < \exp\qty(-\frac{(T-1)}{8}) \leq \exp(-T/16).
\end{align}

\end{proof}

We now combine these two claims and apply them to our setting.
Define the event $[F' \leftarrow G]$ if $F'$ is the set of vertices selected from the graph $G$. Define the distribution $\Pr_{\unif}[\cdot]$ corresponding to first choosing a connected component $C$ with probability proportional to $|C|$, taking $r$ uniform iid samples from the connected component $C$ and setting $F'$ to be the first $|F|$ distinct sampled points. Likewise, define the distribution $\Pr_{\Aa_2}$ corresponding to choosing a random vertex $v$ in $G$, taking $C$ to be the connected component containing $v$, taking $r$ samples according to an expander random walk in $C$ initialized at $v$ with $t$ steps between samples, and then setting $F'$ to be the first $|F|$ distinct sampled points. Set $K$ to be the maximum size of a connected component in $G$ and let $\delta  = (1 + \alpha/2)^t$.  Then we have the following distance bound between the distributions.

\begin{claim}\label{claim:walk-close-to-unif-sets}
Suppose $r, K, \delta$ are such that $rK\delta \leq 10/11$.
For any $F'$ of size $|F'| = |F|$, let $\delta_C(F') = \Pr_{\unif}[F' \leftarrow G] - \Pr_{\Aa_2}[F' \leftarrow G]$. Then 
\begin{equation}
    \abs{\delta_G(F')} \leq \qty(rK\delta + (rK\delta)^2 \frac{1}{1 - rK\delta})  \leq 10 rK\delta.
\end{equation}
Moreover, $\displaystyle \Pr_{\Aa_2}[\mathrm{abort}] \leq 10 rK\delta + \exp(-T/16)$.
\end{claim}
\begin{proof}
For a sequence $v_1,\dots, v_r$ of vertices, let $X_{v_1, \dots, v_r \to F'}$ be the event that $F'$ is the set of the first $|F|$ distinct vertices in $v_1, \dots, v_r$.
\begin{xalign}
    \abs{\delta_G(F')} &= \abs{\sum_{v_1, \dots, v_r} X_{v_1, \dots v_r \to F'} \cdot \qty(\Pr_{\unif}[v_1, \dots, v_r] - \Pr_{\Aa_2}[v_1, \dots, v_r])} \\
    &\leq \sum_{v_1, \dots, v_r} X_{v_1, \dots v_T \to F'} \cdot \abs{\Pr_{\unif}[v_1, \dots, v_r] - \Pr_{\Aa_2}[v_1, \dots, v_r]} \\
    &\leq K^r \cdot \qty(\frac{1}{K})^r \cdot \qty(rK\delta  + (rK\delta)^2 \frac{1}{1 - rK\delta}) \\
    &= \qty(rK\delta + (rK\delta)^2 \frac{1}{1 - rK\delta}) \\ 
    &\leq 10 rK\delta.
\end{xalign}
where we have used \Cref{claim:walk-close-to-unif-points} to bound the difference in probabilities of each sequence of vertices.
Now, to bound the abort probability, let $X_{v_1, \dots, v_r \to \mathrm{abort}}$ be the event that $v_1, \dots, v_r$ contain fewer than $|F|$ distinct vertices.
\begin{xalign}
    \Pr_{\Aa_2}[\mathrm{abort}] &= \sum_{v_1, \dots, v_r} X_{v_1, \dots, v_r \to \mathrm{abort}} \cdot \Pr_{\Aa_2}[v_1, \dots, v_r] \\
    &\leq 10 rK\delta + \sum_{v_1, \dots, v_r} X_{v_1, \dots, v_r \to \mathrm{abort}} \cdot \Pr_{\unif}[v_1, \dots, v_r] \\
    &\leq 10 rK\delta + \exp(-99|F|/8),
\end{xalign}
where we have used \Cref{claim:walk-close-to-unif-points} to replace $\Pr_{\Aa_2}[\cdot]$ by $\Pr_{\unif}[\cdot]$ and then \Cref{claim:unif-doesnt-abort} to bound the abort probability of uniform sampling.
\end{proof}

\subsection{From $\Pr_{\Aa_2}[\cdot]$ to $\Pr_{\ftheng}[\cdot]$}
\noindent We will now proceed to the main argument showing that the pairs $(G,F)$ sampled by $\Aa_2$ are distributed close to the distribution expected by $\Aa_1$. 

\paragraph{$G$ has expanding components with high probability}
To start off, first note that by Lemma \ref{lem:good-expander-whp}, with probability at least $1 - O(N^{-3})$ a graph drawn from $P_{M,\ell}(F)$, for any $F$ of size $\leq N^{1/100}$, will consist of $\ell$ connected-components which are $\alpha$-expanders and have size between $[(1-\gamma)z, (1+\gamma)z]$. Since $\eps_1$ is a constant, for sufficiently large $N$, we can restrict to the situation that the graph is of this form and account for this factor in the end. Henceforth set $K_0 = (1 + \gamma)z$; we are guaranteed that every component has size at most $K_0$.

\paragraph{Relating the probabilities}
In the case that each connected component is an $\alpha$-expander, observe that the probability of every valid pair $(G,F')$ is approximately a constant $p$ independent of $G$ and $F'$. Also recall that the event $F' \leftarrow G$ is the event that $F'$ is the set of vertices selected from $G$.  Moreover, recall the distributions $\mathcal{D}_1$ and $\mathcal{D}_2$ from \Cref{lem:pml-close-to-pmlf}, and notice that $\Dd_2$ is exactly the distribution $\Pr_{\ftheng}$ defined above.
We define
\begin{xalign}
    \delta_{1,2}(G, F') &\defeq \Pr_{\Dd_1}[(G,F')] - \Pr_{\Dd_2}[(G, F')], \\
    \delta_{G}(F') &\defeq \Pr_{\Aa_2}[F' \leftarrow G] - \Pr_{\unif}[F' \leftarrow G].
\end{xalign}
We will now start with the $\Pr_{\Aa_2}$ distribution and bound its distance from $\Pr_{\ftheng}$.
\begin{xalign} 
    \Pr_{\Aa_2}[(G,F')] &= \Pr_{P_{M, \ell}}[G] \cdot \Pr_{\Aa_2}[F' \leftarrow G] \\
    &=\Pr_{P_{M, \ell}}[G ] \cdot \left(\Pr_{\unif}[F' \leftarrow G] + \delta_G(F') \right) = \Pr_{\mathcal{D}_1}[(G,F')] + \delta_G(F') \cdot \Pr_{P_{M,\ell}}[G] \label{eq:choose-f-from-g}  \\
    &= \Pr_{\mathcal{D}_2}[(G,F')] + \delta_{1,2} (G,F') + \delta_{G}(F') \cdot \Pr_{P_{M, \ell}}[G] \\
    &= \Pr_{\ftheng}[(G,F')] + \delta_{1,2} (G,F') + \delta_{G}(F') \cdot \Pr_{P_{M, \ell}}[G].
\end{xalign}

We may now bound the total variational distance between the two sides.
\begin{xalign}
    \norm{ \Pr_{\Aa_2}[\cdot] - \Pr_{\ftheng}[\cdot]} &= \frac{1}{2} \sum_{(G,F')} \abs{\Pr_{\Aa_2}[(G,F')] - \Pr_{\ftheng}[(G,F')]} \\
    &\leq \frac{1}{2} \sum_{(G, F')}\qty( \abs{\delta_{1,2}(G, F')} + \Pr_{P_{M,\ell}}[G] \cdot \abs{\delta_G(F')}) \\
    &\leq \norm{ \Dd_1 - \Dd_2} + \frac{1}{2} \binom{K_0}{|F'|} \cdot \max_{G,F'} \abs{\delta_G(F')} \\
    &\leq 3N^{-9/200} + K_0^{|F|} \cdot (10 r K_0 \delta) \\
    &= 3N^{-9/200}  + ((1+\gamma)z)^{|F| + 1} \cdot (10 \cdot (100 |F|) \cdot (1 - \alpha/2)^t)\\
    &\leq 3N^{-9/200} + \qty( 2N^{9/10})^{N^{1/100} + 1} \cdot 1000 N^{1/100} \cdot ( 1- \alpha/2)^t \\
    &= 3N^{-9/200} + 2000 \cdot 2^{ N^{0.01}} \cdot  N^{0.9 N^{0.01} + 0.91} \cdot (1 - \alpha/2)^t.
\end{xalign}
A total distance bound of $O(N^{-9/200})$ can be achieved if
\begin{xalign}
    2^{N^{0.01}} \cdot N^{0.9 N^{0.01} + 0.91} \cdot (1 - \alpha/2)^t &\leq N^{-9/200}\\
    N^{0.01} + (0.9 N^{0.01} + 0.91) \cdot \log N  + t \cdot \log(1 - \alpha/2) &\leq -\frac{9}{200} \log N \\
   \qty( N^{0.01} \qty(\frac{1}{\log N} + 0.9)+ 0.955  ) \frac{\log N}{\log(1/(1- \alpha/2))} &\leq  t .
\end{xalign}
So setting $t = \Theta(N^{0.02})$ is sufficient. 

Given this choice of $t$, let us know calculate the chance that the sampling of $F'$ aborts. By \Cref{claim:walk-close-to-unif-sets}, this is at most $10 rK\delta + \exp(-T/16) = O(N^{-9/200}) + \exp(-100 N^{0.01}/16) = O(N^{-9/200})$

Thus, the total error probability of $\Aa_2$ equals the error probability of $\Aa_1$ up to 
\begin{equation}
    \underbrace{O(N^{-3})}_{\text{Sample a bad graph}} + \underbrace{O(N^{-9/200})}_{\text{Changing $\Aa_2$ to $\ftheng$}} + \underbrace{O(N^{-9/200})}_{\text{Sampling $F'$ aborts}},
\end{equation} yielding $\eps_2 = \eps_1 - O(N^{-9/200})$ as claimed in theorem.
And the total query complexity assuming not aborting can be calculated as follows. Recall that $t$ was chosen to be $\Theta(N^{0.02})$. The total number of additional queries over $\Aa_1$ is thus the number of steps in the walk which is $100 N^{1/100}  \cdot t \leq O(N^{0.03} ) $. Thus, this algorithm has total query complexity $q + O(N^{0.03})$ and distinguishes with probability $\eps_2$ as claimed.

\section{Wrapping up the proof of Theorem \ref{thm:main}}
\label{sec:putting-it-together}
First, we need to note that the distributions $B_S$ and $P_{M,1}$ which we used as proxies for $\YES$ and $\NO$ instances are not fully supported on $\YES$ and $\NO$ instance graphs, respectively. However, they are very close. For every $S \subset [N]$ of size $\zeta$, let $\widetilde{B}_S$ be the restriction of the distribution $B_S$ (defined in Section \ref{subsec:setup-from-qcma-algorithm}) to graphs with $\ell$ connected components each consisting of between $(1-\gamma)z$ and $(1+\gamma)z$ vertices. By Corollary \ref{cor:QMA-alg-completeness}, the statistical distance between $B_S$ and $\widetilde{B}_S$ is $O(N^{-3})$. 
The $\YES$ instances for Theorem 1 are the  $\{\widetilde{B}_S\}$. 

We consider a single $\NO$ instance of $\widetilde{P}_{M,1}$ where $\widetilde{P}_{M,1}$ is the restriction of $P_{M,1}$ to graphs which are $\alpha$-expanders. The statistical distance between these two distributions is $O(N^{-3})$ by Lemma~\ref{lem:good-expander-whp}.

Furthermore, we can verify that the supports of $\widetilde{P}_{M,1}$ and $\widetilde{B}_S$ are far apart in Hamming distance. Consider graphs $G_1$ and $G_\ell$ from either support, respectively. Consider a connected component $C$ from $G_\ell$. In the graph $G_\ell$, all the edges on $C$ stay within $G_\ell$, but since $G_1$ is an $\alpha$-expander and also a $1/10^2$-edge expander (see proof of Lemma~\ref{lem:good-expander-whp})), then in $G_1$, a $\geq 1/10^4$ fraction of the edges emanating from $C$ leave $C$. As this holds for all components $C$ since $\abs{C} \ll N/2$, then the Hamming distance between the adjacency lists of $G_1$ and $G_\ell$ is $\Omega(N)$. As this holds for all graphs $G_1$ and $G_\ell$, then the Hamming distance bound between the supports hold.

\subsection{$\QMA$ algorithm}

For completeness, from \Cref{cor:QMA-alg-completeness}, we know that the algorithm $\Aa_\QMA$ with witness state $\ket{S}$ answers distribution $\widetilde{B}_S$ with probability at least $\geq 1 - O(N^{-1/20})$. For soundness, from \Cref{lem:good-expander-whp}, we know that $\widetilde{P}_{M,1}$ is an $1/(2 \cdot 10^8)$-expander with probability $\geq 1 - O(N^{-3})$. Therefore, by \Cref{lem:QMA-alg}, the algorithm $\Aa_\QMA$ accepts with probability at most 
\begin{equation}
    \leq 1 - \frac{1}{4} \cdot \frac{1}{(2 \cdot 10^8)} + O(N^{-3}) \leq 1 - \frac{1}{9 \cdot 10^8}.
\end{equation}
By parallel repetition $9 \cdot 10^6 = O(1)$ times, we yield a quantum algorithm with $\leq 0.01$ soundness.

\subsection{$\QCMA$ algorithm}

We argue now that any $\QCMA$ algorithm with completeness $0.99$ and soundness $0.01$ either requires an exponentially long proof or an exponential number of quantum queries. This is done by arguing that any algorithm with a short proof and few queries cannot have such a large completeness and soundness gap. Assume, therefore, that there exists a $\QCMA$ algorithm with a 
\begin{equation}
    q \leq \frac{n \cdot N^{1/100}}{2000} \text{-bit proof and } f = O(N^{1/50}) \text{ quantum queries} 
\end{equation}
and a completeness and soundness gap of $\geq 0.98$. By Lemma \ref{lem:qcma-implies-bqp-for-sunflower}, there exists a 
\begin{equation}
    \qty(\frac{1}{100}, \zeta, \frac{2000 q}{n}) \text{-sunflower $\sunflower$}
\end{equation}
with core $F$ and a $f$-query deterministic quantum algorithm $\Aa$ that accepts each distribution $\widetilde{B}_S$ for $S \in \sunflower$ with probability $\geq 0.99$ and accepts $\widetilde{P}_{M,1}$ with at most $\leq 0.01$ probability. It also accepts that accepts each distribution $B_S$ for $S \in \sunflower$ with probability $\geq 0.99 - O(N^{-3})$ and accepts $P_{M,1}$ with at most $\leq 0.01 + O(N^{-3})$ probability.
Then with the assumed number of queries, we can apply Lemma \ref{lem:distinguishing-sunflowers-and-fixed-distributions} with $\delta = 1/10$ to argue that $\Aa$ must accepts the distribution $H_{\Omega_F}$ with probability $\geq 0.09 - O(N^{-3})$. 
Next, by Lemma \ref{lem:chernoff-tail-bound}, $\Aa$ must accept the distribution $P_{M,\ell}(F)$ with probability $\geq 0.09 - 2 \cdot O(N^{-3}) \geq 0.08$.
We conclude by applying Corollary \ref{cor:poly-method-lb}. Therefore, $\Aa$ must accept the distribution $P_{M,1}$ with probability $>0.02$, a contradiction.
\section{Concluding remarks}

\label{sec:concluding-remarks}

\subsection{Relation to the Fefferman and Kimmel \cite{fefferman-kimmel} construction}

One can think of the result stated in this work as applying the $\QCMA$ lower bounding techniques developed by Fefferman and Kimmel \cite{fefferman-kimmel} to the expander distinguishing problem originally studied by Ambainis, Childs, and Liu \cite{ambainis-childs-liu}.

At a high level, in the \emph{in-place permutation oracle} $\QMA$ and $\QCMA$ separation of \cite{fefferman-kimmel}, the goal was to distinguish between permutations $\fn{\pi}{[N]}{[N]}$ such that $\inv{\pi}([\sqrt{N}])$ is mostly (2/3) supported on odd numbers from permutations mostly supported on even numbers. The original idea in Fefferman and Kimmel was that if the oracle $\pi$ was provided as a classical oracle (an $N n$-bit list $[\pi(1), \pi(2), \ldots, \pi(N)]$) then the subset state $\ket{\xi_{\textrm{ideal}}} = \ket*{\inv{\pi}([\sqrt{N}])}$ would be a good quantum witness.  By measuring the last qubit of a witness $\ket{\xi}$, the verifier can decide if the set $\inv{\pi}([\sqrt{N}])$ is supported mostly on either odd numbers or even numbers. What remains to verify is that the witness $\ket{\xi}$ provided is indeed $\ket{\xi_{\textrm{ideal}}}$. The hope would be to use the oracle for $\pi$ to verify the statement as the sate $\ket*{[\sqrt{N}]}$ can be easily verified by measuring in the Hadamard basis.

However, due to the index-erasure problem, a classical oracle for verifying that $\ket{\xi} =\ket{\xi_{\textrm{ideal}}}$ would need to allow implementation of both $\pi$ and $\inv{\pi}$. However, if the oracle $\inv{\pi}$ is provided, then there is a $\BQP$ algorithm for this problem. Simply, pick a random $j \in [\sqrt{N}]$ and then check if $\inv{\pi}(j)$ is odd or even. The solution in \cite{fefferman-kimmel} was to define the oracle instead as an "in-place oracle'' for $\pi$, meaning a unitary defined as $\sum_j \ketbra{\pi(j)}{j}$. Then the verifier can verify that $\ket{\xi} =\ket{\xi_{\textrm{ideal}}}$ and yet the $\BQP$ algorithm no longer holds.

Fefferman and Kimmel had to make one more modification to prove a $\QMA$ and $\QCMA$ oracle separation: they considered distributions over in-place oracles which mapped to the same ideal quantum witness $\ket{\xi_\textrm{ideal}}$. This was because it seems to be beyond current techniques to prove classical lower bounds without forcing a large structured set of permutations to all share the same witness --- otherwise, for all we know, there might be a mathematical fact about permutations which yields a short classical certificate for any individual permutation. So the oracle is defined as a distribution over unitaries --- i.e. a completely positive trace preserving (CPTP) map.

Notice that this work takes much inspiration form \cite{fefferman-kimmel}; the quantum witnesses for both our work and \cite{fefferman-kimmel} are subset states and we also consider distributions over oracles with the same (or similar) ideal quantum witness. This is because we are unsure how to prove that there is no property of a specific regular graph which yields a short classical witness. We elaborate on why such an impossibility result is hard to prove in the next subsection. What our result principally improves on is that the underlying oracle can be a classical string instead of a unitary.

\subsection{Difficulties in proving stronger statements}

Recall that our $\QMA$ upper bound does not require the setup of distributions over oracles --- it was only included to prove the $\QCMA$ lower bound. How much harder is it (or is it even possible) to prove a $\QCMA$ lower bound without considering distributions?

As pointed out to us by William Kretschmer \cite{kretschmer_2022}, if one considers \emph{average-case} algorithms instead of \emph{worst-case} algorithms, then this problem is $\in \RNP^G$, the average-case analog of $\NP^G$. This is because the average-case version of the expander distinguishing problem is to distinguish the distributions $P_{M,\ell}$ and $P_{M,1}$. And there is a simple randomized algorithm for this problem with a classical witness. Let us recall that for a $d$-regular graph, the expected number of triangles in a connected component is $\Theta(d^3)$ independent of the number of vertices in the component. A similar analysis can be done for $P_{M,\ell}$ and $P_{M,1}$, to show that a random graph from $P_{M,\ell}$ has $\Theta(\ell d^3)$ triangles whereas $P_{M,1}$ has $\Theta(d^3)$ triangles. Therefore, a \emph{classical} witness for the statement that the graph (with high probability) is drawn from $P_{M,\ell}$ (instead of $P_{M,1}$) is a list of $100 \cdot \Theta(d^3)$ triangles from the graph. This witness is easily verifiable and correctly distinguishes with high probability.

Notice that this $\RNP^G$ algorithm does not solve the expander distinguishing problem in the worst-case; since graphs exist in both distributions which are triangle-free (with constant probability). Furthermore, it cannot distinguish the distributions considered in Theorem \ref{thm:main} because the proof relies on finding triangles which is property of the graph not deducible from only knowing the connected components.

But it does highlight a principal roadblock in extending Theorem \ref{thm:main} to distinguishing oracles that are not distributions. It is entirely possible that there exists a property of graphs revealed by looking at the edges that distinguishes graphs with many connected components from graphs with a single expanding connected component. To the best of our knowledge, we do not know of any such property but proving that none exist is beyond the techniques shown here.

Lastly, if we consider the expander distinguishing problem when in the $\YES$ case we are promised that every connected component has size at most $0.99 N$, then this problem is in $\coAM^G$. When the graph is a $\NO$ instance, the verifier can select two random points and the prover can always find a path of length $O(\log N) = O(n)$ between the two. However, when the graph is disconnected and no component is too big, with probability $\geq 1/50$, no path exists.

Therefore, our constructed oracle very finely separates the classes $\QMA$ and $\QCMA$ in the sense that small perturbations of the problem might be very easy.

\onecolumn

\appendix
\section{Omitted concentration inequalities for random graphs}
\label{appendix:chernoff}

\lemgoodexpanderwhp*

\begin{proof}
The proof for $P_{M,1}$ follows as a special case for $\ell = 1$ and $F = \emptyset$.
Consider a graph drawn from the distribution $P_{M,\ell}(F)$ as described in Section \ref{sec:pml-graph-construction}. Let $X_{k,j}$ for $k \in [\ell]$ and $j \in V$ be the indicator random variable that $k_j = i$. Let $X_k = \sum_{j \in V} X_{k,j}$, the size of $\inv{\iota}(V_k)$.
Since the image of a connected component under $\iota$ must lie in some $V_k$, then the size of any connected component is upper-bounded by $(1+\gamma)z$. For any $k > 1$, $\Exp( X_k) = (N - \abs{F})/\ell = z - \abs{F}/\ell$. Since $\abs{F} \ll \ell$, then $\abs{F}/\ell < \gamma z / 2$. Therefore, by a Chernoff bound,
\begin{xalign}
    \Pr\left[\abs{X_k - z} \geq \gamma z\right] &\leq \Pr\left[ \abs{X_k - \Exp (X_k)} \geq \gamma \Exp(X_k) \right] \\
    & \leq 2\exp \left( - \frac{(\gamma/2)^2 \Exp(X_k)}{3} \right) \\
    &\leq 2 \exp \left( - \frac{\gamma^2 N}{12\ell} \right) \\
    & \leq 2 \exp\qty(- \frac{N^{7/10}}{12}). 
\end{xalign}

For $k = 1$, the situation is only slightly different since already $\abs{F}$ terms are guaranteed to be in $X_1$.

A union bound of all these probabilities bounds shows that with all but 
\begin{equation}
2 N^{1/10} \cdot \exp\qty(- \frac{N^{7/10}}{12}) \text{ probability,} \label{eq:connected-components-union-bound}
\end{equation}
the preimage under $\iota$ of every subset $V_k$ has size $\in [(1-\gamma)z,(1+\gamma)z]$.
It remains now to show that, conditioned on this holding, the preimages of every subset $V_k$ are good expanders. Recall the definition of a $\beta$-edge expander is any graph such that for any subset $U \subset V$ of at most half the vertices, $\abs{\partial U} \geq \beta \abs{U}$, where $\partial U$ is the set of neighbors of $U$ excluding $U$ itself.

For any $k \in [\ell]$, let $B_k = \inv{\iota}(V_k)$ be the preimages under $\iota$ of the defined regions of $G'$. If the induced graph on $B_k$ is not a $\beta$-edge expander, there exists some subset $U_1$ of at most half the vertices of $B_k$ and a subset $U_2 \subseteq B_k \setminus U_1$ such that every non-self neighbor of $U_1$ is contained in $U_2$ and, in particular, $\abs{U_2} < \beta \abs{U_1}$. Let $E_{U_1, U_2}$ be the event that this occurs. Note that for a fixed $B_k$, the probability of the event $\Pr[E_{U_1, U_2}] = g(\abs{U_1}, \abs{U_2})$ for some function $g$ --- i.e. the probability only depends on the sizes of the two sets. This is because the graph $G'$ and the injective function $\iota$ are picked independently. Therefore, a union bound on the probability that the connected component is not an edge expander is
\begin{xalign}
    \Pr \left[B_k \text{is \emph{not} a }\beta\text{-edge expander}\right] &\leq \sum_{U_1: \abs{U_1} \leq \abs{B_k}/2} \sum_{U_2 \subseteq V \setminus U_1, \abs{U_2} < \beta \abs{U_1}} \Pr[E_{U_1, U_2}] \\
    &\leq \sum_{i = 1}^{\zeta/2} {\zeta \choose i} {\zeta \choose \beta i}g(i, \beta i).
\end{xalign}
We will soon prove that
\begin{equation}
    g(i, \beta i) \leq \left(2\gamma + \frac{(1+\beta)i}{z}\right)^{\frac{di}{2}}.
    \label{eq:prob-component-k-not-expander}
\end{equation}
Assuming \eqref{eq:prob-component-k-not-expander}, we can bound the probability by using two bounds, one for small $i$ and one for large $i$. For small $i$ --- i.e. whenever $(1+\beta)i \leq 6 \gamma z$, then $g(i, \beta i) \leq \qty(8 \gamma)^{\frac{di}{2}}$ and so
\begin{xalign}
    {\zeta \choose i} {\zeta \choose \beta i} g(i, \beta i) &\leq \zeta^{(1+\beta)i} \cdot \frac{8^{\frac{di}{2}}}{N^{\frac{di}{20}}} \\
    &\leq z^{(1+\beta)i} \cdot 2^{(1+\beta)i} \cdot \frac{8^{\frac{di}{2}}}{N^{\frac{di}{20}}} \\
    &\leq \qty(\frac{2^{1+\beta+3d/2}}{N^{\frac{d}{20} - \frac{9}{10}(1+\beta)}})^i \defeq r^i
\end{xalign}
Assuming $\beta < 4$, then $\frac{d}{20} > \frac{9}{10}(1+\beta)$, and therefore, for sufficiently large $N$, $r \ll \half$. So this geometric series is bounded by
\begin{equation}
\sum_{(1+\beta)i \leq 6 \gamma z}{\zeta \choose i} {\zeta \choose \beta i} g(i, \beta i) \leq \frac{r}{1-r} \leq r(1+2r) \leq 3r \leq N^{-(4-\beta)} \cdot 2^{154}. \label{eq:small-i-sum-all-expander}  
\end{equation}
For large $i$ --- i.e. whenever $(1 + \beta)i > 6\gamma z$ and $i \leq \zeta/2$, note that 
\begin{equation}
g(i, \beta i) \leq \qty( \frac{4}{3} \frac{(1+\beta) i}{z} )^{\frac{di}{2}}.
\end{equation}
Therefore,
\begin{xalign}
    {\zeta \choose i} {\zeta \choose \beta i} g(i, \beta i) &\leq \qty( \frac{e \zeta}{i})^i \qty(\frac{e\zeta}{\beta i})^{\beta i} \qty( \frac{4}{3} \frac{(1+\beta) i}{z} )^{\frac{di}{2}} \\
    &\leq \qty(\frac{(2e)^{1+\beta}}{\beta^\beta} \qty(\frac{4(1+\beta)}{3})^{\frac{d}{2}} \cdot \qty(\frac{i}{z})^{\frac{d}{2} - (1+\beta)})^i \\
    &\leq \qty(\frac{(2e)^{1+\beta}}{\beta^\beta} \qty(\frac{4(1+\beta)}{3})^{\frac{d}{2}} \cdot \qty(\frac{51}{100})^{\frac{d}{2} - (1+\beta)})^i
\end{xalign}
where in the last line we used that $i < \zeta/2$ and $\zeta = (1+\gamma)z$ so for sufficiently large $N$, $i/z \leq 51/100$. For choice of $\beta = 1/100$ and $d = 100$, we get that
\begin{equation}
    {\zeta \choose i} {\zeta \choose \beta i} g(i, \beta i) \leq 2^{-23i}.
\end{equation}
Therefore,
\begin{equation}
\sum_{\frac{6 \gamma z}{1 + \beta} < i < \frac{\zeta}{2}} {\zeta \choose i} {\zeta \choose \beta i} g(i, \beta i) \leq z \cdot 2^{-23 \gamma z} \leq N^{9/10} \cdot 2^{-23 N^{-8/10}}. \label{eq:large-i-sum-all-expander}    
\end{equation}
Then by adding \eqref{eq:small-i-sum-all-expander} and \eqref{eq:large-i-sum-all-expander}, and seeing that clearly \eqref{eq:small-i-sum-all-expander} dominates for sufficiently large $N$, for choice of $\beta = 1/100$,
\begin{equation}
    \Pr[B_k \text{ is \emph{not} a }\beta\text{-edge expander}] \leq 2^{155} N^{-3.99}.
\end{equation}
Consider the graph on $B_k$ induced plus the self-loops introduced. This graph is $d$-regular (including self-loops), then by Cheeger's inequality if it is a $\beta$-edge expander then for $\beta = 1/100$ and $d = 100$, it is a $\alpha$-spectral expander with a second normalized eigenvalue of at most
\begin{equation}
    \alpha \leq 1 - \frac{1}{2 \cdot 10^8}.
\end{equation}
We can now perform one more union bound over both the probability that every $V_k$ is of near-optimal size (\eqref{eq:connected-components-union-bound}) and over all $k \in [\ell]$ that $B_k$ is \emph{not} an $\alpha$-expander. Since $\ell = N^{1/10}$, this overall probability is $O(N^{-3})$ as stated in the statement.

It remains to prove \eqref{eq:prob-component-k-not-expander}. Consider disjoint sets $U_1, U_2$ contained in $B_k$
where $\abs{U_1} = i$ and $\abs{U_2} = \beta i$. Then, the event $E_{U_1, U_2}$ is equivalent to the event that every edge emanating from $U_1$ is contained in $U_1 \cup U_2$. Since the graph $G$ from $P_{M,\ell}(F)$ is built by considering a graph $G'$ on $M$ vertices built of $d$ perfect matchings and then considering the induced graph on $N$ vertices under the injective map $\fn{\iota}{V}{V'}$, then this implies that every edge emanating from $\iota(U_1)$ is contained in
\begin{equation}
 A \defeq \qty(V_k \setminus \iota(B_k)) \cup \iota(U_1) \cup \iota(U_2).
\end{equation}
the size of $A$ is easily bounded as $\abs{A} \leq [z(1 + \gamma) - z(1-\gamma)] + i + \beta i = 2\gamma z + (1+\beta)i$.
For each of the $d$ matchings (corresponding to a different color), the probability that all $\kappa$-th colored edges emanating $\iota(U_1)$ lie in $A$ is
\begin{equation}
 \frac{\abs{A} - 1}{\zeta - 1} \cdot \frac{\abs{A} - 3}{\zeta-3} \cdot \ldots \cdot \frac{\abs{A} - i + 1}{\zeta - i + 1} \leq \qty(\frac{\abs{A}}{\zeta})^{i/2} \leq \qty(\frac{\abs{A}}{z})^{i/2} \leq \qty(2 \gamma + \frac{(1+\beta)i}{z})^{i/2}.
\end{equation}
Since the $d$ edge colorings are independently sampled, then the net probability is bounded by \eqref{eq:prob-component-k-not-expander}.
 
\end{proof}

\lempmlclosetopmlf*

\begin{proof}
In order to prove this lemma, it is illustrative to decompose the procedure for sampling the distributions $P_{M,\ell}$ and $P_{M,\ell}(F)$. Let us note that while the procedure is stated sequentially, there are three different independent components. First, the sampling of the graph $G'$ on $M$ vertices is independent of the construction of the injective map $\iota$. Furthermore sampling $\iota$ is constructed by independently sampling $\fn{k}{V}{[\ell]}$ and then defining the injective map $\iota: V \hookrightarrow V'$ by assigning each vertex $j$ to a vertex in $V_{k(j)}$ without replacement. An equivalent sampling algorithm is to sample uniformly random independent injective maps $\pi_1, \ldots, \pi_\ell$ where each $\pi_i: [M/\ell] \hookrightarrow V_i$ is an enumeration of the vertices of $V_i$. Then, define $\iota$ by using the random enumerations $\{\pi_i\}$ to sequentially assign each vertex $j$ a vertex in $V_{k(j)}$. Formally, if $j$ is the $s$-th vertex such that $k(j) = i$; then $\iota(j) = \pi_i(s)$.

Therefore, the sampling procedure can be rewritten as a deterministic process based on the three samples: the graph $G'$, the map $k$, and enumerations $\{\pi_i\}$. Notice that the only difference between $P_{M,\ell}$ and $P_{M,\ell}(F)$ is the map $k$. Notice, we can further simplify by thinking only of the map $\fn{k'}{V}{\bits}$ where $k' = 0$ if $k \neq 1$. This is because we can sample a uniformly random function $\fn{k''}{V}{[\ell] \setminus \{1\}}$ and define
\begin{equation}
    k(j) = \begin{cases} 1 & \text{if } k'(j) = 1 \\
    k''(j) & \text{if } k'(j) \neq 1.
    \end{cases}
\end{equation}

{Let us say that $k': V \to \{0,1\}$ is a \emph{uniformly random map} if each $k(j)$ for $j \in V$ is an iid random variable with $\Pr[k'(j) = 1] = 1/\ell$.}
Now, let $\Dd_1'$ be the distribution on $(k',F)$ defined by sampling a uniformly random map $k'$ and then uniformly randomly a set $F$ of size $m$ from $\inv{(k')}(1)$. Let $\Dd_2'$ be the distribution on $(k,F)$ formed by sampling a uniformly random set $F$ of $m$ vertices and then a uniformly random map $k'$ such that $k'(F) = 1$. By the previously stated decomposition of the sampling procedures for $\Dd_1$ and $\Dd_2$, if we show that $\Dd_1'$ and $\Dd_2'$ are statistically indistinguishable, then $\Dd_1$ and $\Dd_2$ are at least as indistinguishable. 

We first remark that these distributions are not the same as the expected Hamming weight of a vector $k'$ from $\Dd_1'$ is $N/\ell$ whereas the expected Hamming weight of the vector from $\Dd_2'$ is $m + (N-m)/\ell$. To show, however, that these distributions are statistically close, we will use some simple Chernoff bounds and Pinsker's inequality. Let $E$ be the event that the sampled vector $k'$ has expected Hamming weight either $< (1-\gamma) z$ or $> (1+\gamma) z$. Let $a_1$ and $a_2$ be the probabilities of the event $E$ over distributions $\Dd_1'$ and $\Dd_2'$. Since $m \ll z$, a simple Chernoff bound shows that both $a_1$ and $a_2$ are at most
\begin{equation}
    a_1, a_2 \leq 2 \exp \qty(-\frac{\gamma^2 N}{12 \ell}) \leq 2 \exp \qty(-\frac{N^{7/10}}{12}).
\end{equation}
Let $\Dd_1''$ and $\Dd_2''$ be the distributions conditioned on the event $\lnot E$, respectively. Therefore, the statistical distances are at most
\begin{equation}
    \norm{\Dd_1' - \Dd_1''} \leq 2a_1, \quad \norm{\Dd_2' - \Dd_2''} \leq 2 a_2. \label{eq:stat-distance}
\end{equation}
We now bound the statistical distance between $\Dd_1''$ and $\Dd_2''$ using Pinsker's inequality and a bound on the KL divergence. We calculate the probability of outputting $(k', F)$ under both distributions. First,
\begin{equation}
    \Dd_1''(k', F) = \frac{1}{1-a_1} \cdot \qty(\frac{1}{\ell})^{\abs{k'}} \qty(1 - \frac{1}{\ell})^{N - \abs{k'}} \cdot \frac{1}{{\abs{k'} \choose m}}.
\end{equation}
The $\inv{(1-a_1)}$ term is due to the conditioning on event $\lnot E$. The next two terms give the probability of sampling $k'$ as each index is set to $1$ with iid probability $1/\ell$. The last is the probability of selecting the specific set $F$ given $k'$. Second,
\begin{equation}
    \Dd_2''(k', F) = \frac{1}{1-a_2} \cdot \frac{1}{{N \choose m}} \cdot \qty(\frac{1}{\ell})^{\abs{k'}-m} \qty(1 -\frac{1}{\ell})^{(N-m) - (\abs{k'}-m)}.
\end{equation}
Likewise, the $\inv{(1-a_2)}$ term is due to the conditioning on event $\lnot E$. The next term is the probability of sampling $F$. The last two terms yield the probability of sampling the rest of $k'$.
Then, as a sub-calculation in the KL divergence,
\begin{xalign}
\ln \qty(\frac{\Dd_1''(k',F)}{\Dd_2''(k',F)}) &= \ln \qty( \frac{1-a_2}{1-a_1} \cdot \qty(\frac{1}{\ell})^m \cdot \frac{{N \choose m}}{{\abs{k'} \choose m}}) \\
&= \ln \qty( \frac{1-a_2}{1-a_1} \cdot \frac{N}{\abs{k'} \ell} \cdot \frac{N-1}{(\abs{k'}-1)\ell} \cdot \ldots \cdot \frac{N - m + 1}{(\abs{k'}-m+1)\ell}) \\
&\leq \ln(\frac{1-a_2}{1-a_1}) + \sum_{i = 0}^{m-1} \ln(\frac{N - i}{(\abs{k'}-i)\ell}) \\
&\leq 2 a_1 + \sum_{i = 0}^{m-1} \ln(\frac{N - i}{((1-\gamma) z -i)} \cdot \frac{z}{N}) \label{eq:KL-calc1} \\
&\leq 2 a_1 + \sum_{i = 0}^{m-1} \ln(\frac{z}{(1-\gamma) z -i}) \\
&\leq 2 a_1 + \sum_{i = 0}^{m-1} \ln(\frac{1}{1-2\gamma}) \label{eq:KL-calc2} \\
&\leq 2 a_1 + 4 m \gamma \label{eq:KL-calc3} \\
&\leq 5 N^{-9/100} \label{eq:KL-calc4}.
\end{xalign}
Here we used $\ln(\frac{1}{1-x}) \leq 2x$ twice and \eqref{eq:KL-calc1} follows from $\abs{k'} \geq (1-\gamma)z$ since we condition on the event $\lnot E$ and \eqref{eq:KL-calc2} follows from $m \ll \gamma z$. To pass from \eqref{eq:KL-calc3} to \eqref{eq:KL-calc4}, we used the fact that $2\alpha_1 \ll m\gamma \leq N^{-9/100}$.
Therefore, the total KL divergence between $\Dd_1''$ and $\Dd_2''$ is also bounded by 
\begin{equation}
    \textrm{KL}(\Dd_1'', \Dd_2'') = \sum_{(k',F)} \Dd_1''(k', F) \ln\qty(\frac{\Dd_1''(k',F)}{\Dd_2''(k',F)}) \leq 5 N^{-9/100} \sum_{(k', F)} \Dd_1''(k', F) = 5N^{-9/100}.
\end{equation}
 By Pinsker's inequality, then
\begin{equation}
    \norm{\Dd_1'' - \Dd_2''} \leq \sqrt{\half \cdot \textrm{KL}(\Dd_1'', \Dd_2'')} \leq \sqrt{\half \cdot 5 N^{-9/100}} \leq 2 N^{-9/200}.
\end{equation}
As this dominates the bounds in \eqref{eq:stat-distance}, we can conclude by the triangle inequality for $\norm{\cdot}$ that for sufficiently large $N$, $\norm{\Dd_1' - \Dd_2'} \leq 3 N^{-9/200}$. Finally, as remarked previously, this implies that $\norm{\Dd_1 - \Dd_2} \leq 3 N^{-9/200}$.
\end{proof}

\section{Ommited proofs for the adversary method}

\label{appendix:adversary-proofs}

Every oracle query algorithm using Hilbert space $\Hh_A$ can be adapted into a "ficticious'' oracle query algorithm using Hilbert space $\Hh_A \otimes \Hh_I$ where an oracle gate $\ket{i} \mapsto (-1)^{a_i} \ket{i}$ is turned into a gate $\ket{i}\ket{a,b} \mapsto (-1)^{a_i} \ket{i}\ket{a,b}$ where the second register is $\Hh_I$. this allows us to analyze how the algorithm would behave on a superposition over oracles.

Assume an algorithm $\Aa$ makes $T$ queries to the oracle. For $1 \leq t \leq T$, let $\rho_t = \tr_{\Hh_A}(\ketbra{\psi_t})$ be the reduced density matrix (for corresponding pure state $\ket{\psi_t}$) of the computation on the register $\Hh_I$ immediately before the $t$-th oracle gate. We assume the algorithm from an initial starting state of $\ket{0}_{\Hh_A} \otimes \ket{\chi_0}_{\Hh_I}$ evolves to
\begin{equation}
    \ket{\psi_0} = \ket{\xi_{0}}_{\Hh_A} \otimes \ket{\chi_{0}}_{\Hh_I}
\end{equation}
immediately before the application of the first oracle query.
The convenient start, as suggested by Ambainis, is to consider the algorithm run with
\begin{equation}
    \ket{\chi_{0}} = \frac{1}{\sqrt{2\abs{X}}} \sum_{x \in X} \ket{x} + \frac{1}{\sqrt{2\abs{Y}}} \sum_{y\in Y} \ket{y}.
\end{equation}
In this case\footnote{Ambainis uses the notation $\rho_{xy} = \bra{y}\rho \ket{x}$ to refer to indices of matrices (this matches the standard notation).}, for $x \in X$ and $y \in Y$,
\begin{equation}
    (\rho_0)_{xy} = \frac{1}{2 \sqrt{\abs{X}\abs{Y}}}.
\end{equation}
For any algorithm $\Aa$ achieving \eqref{eq:alg-assumption}, let $X_{\good}$ and $Y_{\good}$ be the set of $x$ and $y$, respectively, such that $\Pr_\Aa[\Aa^x = 0]$ and $\Pr_\Aa[\Aa^y = 1]$ are $\geq 1 - \eps$. By Markov's inequality, $\abs{X_\good} \geq (1 - \eps) \abs{X}$ and $\abs{Y_\good} \geq (1 - \eps) \abs{Y}$. For any $(x,y) \in X_\good \times Y_\good$, we will show that
\begin{equation}
    \abs{\qty(\rho_T)_{xy}} \leq \sqrt{\frac{\eps(1-\eps)}{\abs{X}\abs{Y}}}. \label{eq:correctly-sorted}
\end{equation}
To show \eqref{eq:correctly-sorted}, let $\ket*{\psi_x}$ and $\ket*{\psi_y}$ be the final states of $\Aa$ when run with inputs $\ket{x}$ and $\ket{y}$ in register $\Hh_I$, respectively. Then, the final state of the algorithm after $T$ queries will be
\begin{equation}
    \ket{\psi_{\text{final}}} = \frac{1}{\sqrt{2\abs{X}}} \sum_{x \in X} \ket*{\psi_x} \ket{x} + \frac{1}{\sqrt{2\abs{Y}}} \sum_{y\in Y} \ket*{\psi_y}\ket{y}.
\end{equation}
Take a basis of $\Hh_A$ of the form $\ket{z}\ket{v}$ where $\ket{z}$ corresponds to the answer bit and $\ket{v}$ is a basis for the remainder of the work register. In this basis, let
\begin{equation}
    \ket*{\psi_x} = \sum_{z, v} \alpha_{z,v} \ket{z}\ket{v}, \qquad \ket*{\psi_y} = \sum_{z,v} \beta_{z,v} \ket{z}\ket{v}.
\end{equation}
Since the algorithm $\Aa$ cannot effect the amplitudes of the oracle string in $\Hh_A$, then 
\begin{xalign}
    (\rho_T)_{xy} &= \bra{y} \rho_T \ket{x} \\
    &= \sum_{z, v} \ip{z,v,y}{\psi_{\text{final}}}\ip{\psi_{\text{final}}}{z,v,x} \\
    &= \sum_{z,v} \frac{\alpha_{z,v}^\dagger}{\sqrt{2\abs{Y}}} \frac{\beta_{z,v}}{\sqrt{2\abs{X}}} \\
    &= \frac{1}{2 \sqrt{\abs{X}\abs{Y}}} \sum_{z,v} \alpha_{z,v}^\dagger \beta_{z,v}. \label{eq:rho-T-bound}
\end{xalign}
Like Ambainis, define
\begin{equation}
    \eps_0 \defeq \sum_{v} \abs{\alpha_{0,v}}^2, \qquad \eps_1 \defeq \sum_{v} \abs{\beta_{1,v}}^2.
\end{equation}
Since $x \in X_\good$ and $y \in Y_\good$, then $\eps_0 \leq \eps$ and $\eps_1 \leq \eps$. Then
\begin{xalign}
    \sum_{z,v} \alpha_{z,v}^\dagger \beta_{z,v} &\leq \sum_{z,v} \abs{\alpha_{z,v}}{\abs{\beta_{z,v}}} \\
    &= \sum_{v} \abs{\alpha_{0,v}}{\abs{\beta_{0,v}}} + \abs{\alpha_{1,v}}{\abs{\beta_{1,v}}} \\
    &\leq \sqrt{\sum_{v} \abs{\alpha_{0,v}}^2 } \sqrt{\sum_{v} \abs{\beta_{0,v}}^2 } + \sqrt{\sum_{v} \abs{\alpha_{1,v}}^2 } \sqrt{\sum_{v} \abs{\beta_{1,v}}^2 } \label{eq:lem-output-cauchy} \\
    &= \sqrt{\eps_0\qty(1 - \eps_1)} + \sqrt{\eps_1\qty(1-\eps_0)} \\
    &\leq 2 \sqrt{ \eps\qty(1-\eps)}. \label{eq:lem-output-ip-bound}
\end{xalign}
where \eqref{eq:lem-output-cauchy} follows from the Cauchy-Schwarz inequality. Combining \eqref{eq:rho-T-bound} and \eqref{eq:lem-output-ip-bound} gives \eqref{eq:correctly-sorted}.

\thmambainisextension*

\begin{proof}
Let $X_\good$ and $Y_\good$ be the sets as previously defined and let
\begin{equation}
    R_\good \defeq R \cap X_\good \times Y_\good.
\end{equation}
We can bound the size of $R_\good$ as follows. First note that
\begin{xalign}
    \abs{R} &\geq \max \qty(\underline{m} \abs{X}, \underline{m'} \abs{Y}) \\
    &\geq \frac{\underline{m} \abs{X} + \underline{m'} \abs{Y}}{2}. 
\end{xalign}
Second, by the upper bound on the degree,
\begin{xalign}
    \abs{R \setminus R_\good} &\leq \overline{m} \abs{X \setminus X_\good} + \overline{m'} \abs{Y \setminus Y_\good} \\
    &\leq \eps \overline{m} \abs{X} + \eps \overline{m'} \abs{Y}.
\end{xalign}
Therefore,
\begin{xalign}
    \abs{R_\good} \geq \half \qty( \qty(\underline{m} - 2 \eps \overline{m}) \abs{X} + \qty(\underline{m'} - 2 \eps \overline{m'}) \abs{Y}) \\
    \geq \sqrt{\qty(\underline{m} - 2 \eps \overline{m})  \qty(\underline{m'} - 2 \eps \overline{m'})}\cdot \sqrt{\abs{X} \abs{Y}.}
\end{xalign}
Now define,
\begin{equation}
    S_t \defeq \sum_{(x,y) \in R_\good} \abs{(\rho_t)_{xy}}.
\end{equation}
Notice that then
\begin{xalign}
    S_0 - S_T &= \sum_{(x,y) \in R_\good} \abs{(\rho_0)_{xy}} - \abs{(\rho_T)_{xy}} \\
    &\geq \abs{R_\good} \cdot \qty(\frac{1 - 2\sqrt{\eps(1-\eps)}}{2 \sqrt{\abs{X}\abs{Y}}}) \\
    &\geq 
    \sqrt{\qty(\underline{m} - 2 \eps \overline{m})  \qty(\underline{m'} - 2 \eps \overline{m'})} \cdot \frac{1 - 2 \sqrt{\eps(1-\eps)}}{2}.
\end{xalign}
We will next show that the difference $S_t - S_{t+1}$ is minimal:
\begin{equation}
    S_t - S_{t+1} \leq \frac{\sqrt{\ell_{\max}}}{2}. \label{eq:difference-per-query}
\end{equation}
Therefore,
\begin{equation}
    S_0 - S_T \leq \frac{T \sqrt{\ell_{\max}}}{2}
\end{equation}
which yields a lower bound of
\begin{equation}
    T \geq \qty(1 - 2 \sqrt{\eps(1-\eps)}) \sqrt{\frac{\qty(\underline{m} - 2 \eps \overline{m})  \qty(\underline{m'} - 2 \eps \overline{m'})}{\ell_{\max}}}.
\end{equation}
To show \eqref{eq:difference-per-query}, we first write the intermediate state $\ket{\psi_t}$ as
\begin{equation}
    \ket{\psi_t} = \sum_{i,r, z} \alpha_{i,r,x,z} \ket{i, r, z} \otimes \ket{x}
\end{equation}
where $i \in [N]$ is the index of the input variable being queried, $r$ is the answer bit for the query, $z$ are the bits not involved in the query or answer, and $x$ is the oracle on $N + M$ bits being queried. Immediately after querying the oracle, the state of the algorithm is precisely
\begin{equation}
    \sum_{i,r,z} \alpha_{i,r,x,z} \ket{i, r \oplus x_i, z} \otimes \ket{x}.
\end{equation}
Like Ambainis, if we denote
\begin{xalign}
\ket{\xi_{i,r,z}} &= \sum_x \alpha_{i,r,x,z} \ket{x}, \\
\ket{\xi_{i,r,z}'} &= \sum_x \alpha_{i,r\oplus x_i, x, z} \ket{x}, \\
\rho_{t,i} &= \sum_{r,z} \ketbra{\xi_{i,r,z}}, \\
\text{ and } \rho_{t+1,i} &= \sum_{r,z} \ketbra{\xi_{i,r,z}'},
\end{xalign}
then, notice $\rho_{t,i}$ and $\rho_{t+1,i}$ are the parts of $\rho_t$ and $\rho_{t+1}$ corresponding to querying index $i$, with each being the sum over $i \in [N]$ of the corresponding parts.
Notice that
\begin{xalign}
    \abs{(\rho_t)_{xy}} - \abs{(\rho_{t+1})_{xy}} &\leq \abs{(\rho_t)_{xy} - (\rho_{t+1})_{xy}} \\
    &= \abs{\sum_i (\rho_{t,i})_{xy} - (\rho_{t+1,i})_{xy}} \\
    &\leq \sum_i \abs{(\rho_{t,i})_{xy} - (\rho_{t+1,i})_{xy}}. \label{eq:sum-over-small-i}
\end{xalign}
Notice that if $x_i = y_i$, then
\begin{equation}
    (\rho_{t,i})_{xy} = \sum_{r,z} \alpha_{i,r,x,z}^\dagger \alpha_{i,r,y,z} = \sum_{r,z} \alpha_{i,r \oplus x_i, x, z}^\dagger \alpha_{i,r \oplus y_i, y, z} = (\rho_{t+1,i})_{xy}.
\end{equation}
And when $x_i \neq y_i$, then
\begin{equation}
    \qty(\rho_{t+1,i})_{xy} = - \qty(\rho_{t,i})_{xy} \implies \abs{\qty(\rho_{t,i})_{xy} - \qty(\rho_{t+1,i})_{xy}} = 2 \abs{\qty(\rho_{t,i})_{xy}}.
\end{equation}
Therefore, the only indices we need to consider in \eqref{eq:sum-over-small-i} are $i \in [N]$ for which $x_i \neq y_i$. 
Let $S_{t,i}$ equal
\begin{equation}
    S_{t,i} \defeq \sum_{(x,y) \in R_\good \st x_i \neq y_i } \abs{(\rho_{t,i})_{xy}}.
\end{equation}
Then,
\begin{equation}
    S_t - S_{t+1} = \sum_{(x,y) \in R_\good} \abs{(\rho_t)_{xy}} - \abs{(\rho_{t+1})_{xy}} \leq 2 \sum_{i \in [N]} S_{t,i}.
\end{equation}
To bound $S_{t,i}$, since $\rho_{t,i}$ is non-negative by construction, 
\begin{xalign}
    0 &\leq \qty(\bra{y}\sqrt{\ell_{x,i}} - \bra{x}\sqrt{\ell_{y,i}}) \rho_{t,i} \qty(\sqrt{\ell_{x,i}}\ket{y} - \sqrt{\ell_{y,i}} \ket{x}) \\
    &= \ell_{x,i} \qty(\rho_{t,i})_{yy} + \ell_{x,i}\qty(\rho_{t,i})_{xx} - \sqrt{\ell_{x,i}\ell_{y,i}} \qty( \qty(\rho_{t,i})_{xy} + \qty(\rho_{t,i})_{yx})
\end{xalign}
and therefore
\begin{equation}
    \abs{ \qty(\rho_{t,i})_{xy}} \leq \half \qty( \sqrt{\frac{\ell_{y,i}}{\ell_{x,i}}} \abs{\qty(\rho_{t,i})_{xx}} + \sqrt{\frac{\ell_{x,i}}{\ell_{y,i}}} \abs{\qty(\rho_{t,i})_{yy}}).
\end{equation}
So, 
\begin{xalign}
    S_{t,i} &\leq \sum_{(x,y) \in R_\good \st x_i \neq y_i } \half \qty( \sqrt{\frac{\ell_{y,i}}{\ell_{x,i}}} \abs{\qty(\rho_{t,i})_{xx}} + \sqrt{\frac{\ell_{x,i}}{\ell_{y,i}}} \abs{\qty(\rho_{t,i})_{yy}}) \\
    &\leq \half \qty(\sum_{x \in X_\good} \sqrt{\abs{\ell_{x,i}\ell_{y,i}}} \abs{\qty(\rho_{t,i})_{xx}} + \sum_{y \in Y_\good} \sqrt{\abs{\ell_{x,i}\ell_{y,i}}} \abs{\qty(\rho_{t,i})_{yy}} ) \\
    &\leq \frac{\sqrt{\ell_{\max}}}{2} \tr(\rho_{t,i}). \label{eq:follows-from-positivity}
\end{xalign}
where \eqref{eq:follows-from-positivity} follows from positivity and the definition of trace.
Therefore,
\begin{equation}
    S_t - S_{t+1} = \sum_{i \in [N]} S_{t,i} \leq \frac{\sqrt{\ell_{\max}}}{2} \sum_{i \in [N]} \tr(\rho_{t,i}) = \frac{\sqrt{\ell_{\max}}}{2} \tr(\rho_{t}) = \frac{\sqrt{\ell_{\max}}}{2}.
\end{equation}

\end{proof}

\corambainisextension*

\begin{proof}
Suppose there exists an algorithm $\Aa$ such that
\begin{equation}
    \Exp_{y \in Y} \Pr_{\Aa}[\Aa^y = 1] - \Exp_{x \in X} \Pr_{\Aa}[\Aa^x = 0] \geq 1 - \delta.
\end{equation}
Then,
\begin{equation}
    \Exp_{y \in Y} \Pr_{\Aa}[\Aa^y = 1] \geq 1 - 2\delta \quad \text{and} \quad \Exp_{x \in X} \Pr_{\Aa}[\Aa^x = 0] \leq 2\delta.
\end{equation}
Define $X' = X \times \{0\}$ and $Y' = Y \times \{1\}$. Then, these sets in $\bits^{N+M+1}$ are disjoint and satisfy \eqref{eq:alg-assumption} for $\eps = 2\delta$ and we can apply Theorem~\ref{thm:ambainis-extension}.
\end{proof}

\end{document}